\documentclass[a4paper,12pt]{article}
\usepackage{epsfig}
\usepackage{graphicx}
\usepackage{amsmath}
\usepackage{amsfonts}
\usepackage{dsfont}
\usepackage{amssymb}
\usepackage{mathrsfs}
\usepackage{subfigure}
\usepackage{wasysym}
\newcommand{\be}{\begin{equation}}
\newcommand{\ee}{\end{equation}}
\newcommand{\bea}{\begin{eqnarray}}
\newcommand{\eea}{\end{eqnarray}}

\newcommand{\p}{\textrm{{\scriptsize(+)}}}
\newcommand{\m}{\textrm{{\scriptsize(--)}}}

\newcommand{\tpm}{(\pm)}

\newcommand{\im}{\textrm{Im}}
\newcommand{\re}{\textrm{Re}}

\DeclareMathOperator{\cn}{cn}

\DeclareMathOperator{\e}{e}
\title{\textbf{Non-adiabatic quantum effects from a Standard Model
    time-dependent Higgs \emph{vev}}}
\author{ 
  \Large{R.~Casadio\footnotemark[1]~,
         P.~L.~Iafelice\footnotemark[2]~ and 
         G.~P.~Vacca\footnotemark[3]}\\[6pt]
  \textit{Dip.~di Fisica - Universit\`a di
             Bologna and INFN - Sezione di Bologna,}\\[0pt]
  \textit{via Irnerio 46, 40126 Bologna, Italy}}
\date{\today}
\begin{document}
\footnotetext[1]{email: casadio@bo.infn.it}
\footnotetext[2]{email: iafelice@bo.infn.it}
\footnotetext[3]{email: vacca@bo.infn.it}
\maketitle
\begin{abstract}
\noindent
We consider the time-dependence of the Higgs vacuum expectation value
(\emph{vev}) given by the dynamics of the Standard Model and study the
non-adiabatic production of both bosons and fermions,
which is intrinsically non-perturbative.
In the Hartree approximation, we analyse the general expressions that describe
the dissipative dynamics due to the back-reaction of the produced particles.
In particular, we solve numerically some relevant cases for the Standard
Model phenomenology in the regime of relatively small oscillations of the
Higgs~\emph{vev}.
\end{abstract}
\section{Introduction}
\label{intro}
\setcounter{equation}{0}
In the Standard Model, several fundamental constants such as the Fermi
coupling $G$ and the mass of gauge bosons and fermions depend on the
vacuum expectation value (\emph{vev}) of the Higgs field, because of the
well-known mechanism of spontaneous symmetry breaking.
The equation of motion of the Higgs field on the other hand allows for
(periodic) time-dependent solutions for the Higgs \emph{vev}, which can
then be viewed as a particular case of varying fundamental constants
(for a review, see Ref.~\cite{uzan}).
This behaviour differs from the more common cases of adiabatic variations
of the fundamental constants and, due to its periodic nature, can lead to
efficient (non-adiabatic) particle production. 
\par
The issue of the constancy of
the constants of physics was (probably) first addressed by P.~A.~M.~Dirac
\cite{dirac-old,dirac-new} with his ``Large Numbers hypothesis'': very large (or small)
dimensionless universal constants cannot be pure mathematical numbers and must
not occur in the basic laws of physics.
He then proposed that they be considered as (typically slowly) variable parameters
characterising the state of the Universe and pointed out the possibility of measuring
astrophysical quantities to settle this question ~\cite{kujat,hannestad,kaplinghat}. 
More recently, theories such as string theory and models with extra spatial
dimensions have also predicted the time-dependence of the phenomenological
constants of the low energy regime describing our Universe \cite{Taylor-Veneziano,Witten}.
\par
The cases we shall consider instead correspond to oscillations of the Higgs
\emph{vev} with periods (set by the Higgs mass scale) of the order of
$10^{-26}\,$s.
Such a behaviour can be obtained from the usual dynamics of the spatially
homogeneous Higgs field~\cite{passarino}.
We therefore begin by considering the classical equation of motion for the
time-dependent Higgs \emph{vev} (i.e. a classical condensate) in a homogeneous
patch of space-time and identifying the relevant regimes.
Quantum fluctuations of fermion and boson fields of the Standard
Model are then analysed on this Higgs background and explicit
expressions for the particle production~\cite{prod}, an intrinsically non-perturbative
effect, are presented.
In particular, we investigate which bosons and fermions are produced
more abundantly depending on the Higgs mass which we consider
in a physically sound range of values~\cite{higgs-mass}. 
Further, the back-reaction of particle production~\cite{ba} is analysed
in the Hartree approximation which is well suited to describe
such a dissipative effect.
Similar methods have been previously employed to study particle production
in strong fields~\cite{Mostepanenko_book} and pre- as well as re-heating
in Cosmology~\cite{Boyan-deVega,kof-linde-starob,heat}.
\par
According to our findings, particle production induced by the oscillating
Higgs can be very efficient.
From the phenomenological point of view, if the Higgs were oscillating now,
the amplitudes of such oscillations should therefore be extremely small.
On the other hand, this mechanism could explain how Higgs kinetic and
potential energy have dissipated in the past~\cite{Boyan-deVega}.
Note that we typically consider regimes such that only small oscillations around
an absolute minimum are present and the symmetry breaking phase transition
is not significantly affected.
However, since we solve the complete system of coupled non-linear equations
that describe particle production and their back-reaction, one does not expect
that the produced particles are thermal~\cite{boya2}.
They also turn out to have mostly small momenta and are therefore non-relativistic.
\par
In Section~\ref{higgs_t}, we review the solutions of the classical equation of
motion for the Higgs \emph{vev} to properly identify the periodicity properties.
Fermion modes on such a Higgs background are then
studied in Section~\ref{sec-fermion_theory}, with a particular emphasis on the
introduction of physical quantities and the regions which lead to quantum particle
production.
The same kind of analysis is then given for the bosons in Section~\ref{analytic-boson}.
Fermion and boson production within the Standard Model are then studied in
details in Sections~\ref{fermi_pro} and
\ref{boso_pro}, respectively, taking into account the dependence on the Higgs
mass. At this stage we neglect the back-reaction.
This latter effect, which has to be included to describe consistently the
dynamics is then considered in Section~\ref{back}. In this section we analyze phenomena
like the dissipation of the Higgs energy and particle production, consistent
with total energy conservation. 
Final comments are given in Section~\ref{conc}.
\par
We shall use natural units with $\hbar=c=1$.
%
%
%
\section{Higgs \emph{vev} time dependence}
\label{higgs_t}
\setcounter{equation}{0}
The Higgs sector of the Standard Model Lagrangian we are interested in
is given by \cite{SMpassarino}:
\begin{equation}
\label{lagrH}
{\mathcal L}_{H} = \partial^{\mu} H^{\dagger} \partial_{\mu}H - {\mu}^2\,H^{\dagger}H -
                  \frac{1}{2}\,\lambda\,( H^{\dagger}H)^2,
\end{equation}
where $H$ is the complex iso-doublet
\begin{equation}
H = \frac{1}{\sqrt{2}}
\exp \left( i\frac{\xi^{a}\tau^{a}}{v}\right)
\left[ v + h(x) \right]
\binom{0}{1},
\end{equation}
$\tau^{a}$ (with $a=1,2,3$) are the $SU(2)_{L}$ generators, $\xi^{a}$ and $h$
(with $\langle0|\xi^{a}|0\rangle =0= \langle0|h|0\rangle$)
are scalar fields which parameterise the fluctuations of $H$ around the vacuum state.
We have neglected the couplings with gauge and fermion fields.
\par
We find it useful to introduce two new quantities, $M_{H}$ and $\beta$,
defined as
\begin{equation}
  M_{H} = 2\,\frac{M}{g}\sqrt{\lambda},
  \qquad 
  \mu^2 = \beta - \frac{1}{2}\,M_{H}^{2},
\end{equation}
where $g$ is the $SU(2)_{L}$ coupling constant and $M_{H}$ is the ``bare'' Higgs
boson mass (corresponding to constant Higgs \emph{vev}).
In the same way, $M$ is the ``bare'' weak vector boson mass. 
The parameter $\beta$ is such that $\langle0|h|0\rangle =0$ to all orders in
perturbation theory.
Since we are going to neglet the loop corrections, we set $\beta=0$ hereafter.
Note also that the potential in the Lagrangian \eqref{lagrH} yields a partial spontaneous
symmetry breaking (SSB).
In fact, ground-state configurations are only invariant under
$U(1)_{em} \subset SU(2)_{L}\otimes U(1)_{Y}$.
\par
\begin{figure}[t]
   \begin{center}
     \includegraphics[scale=0.9]{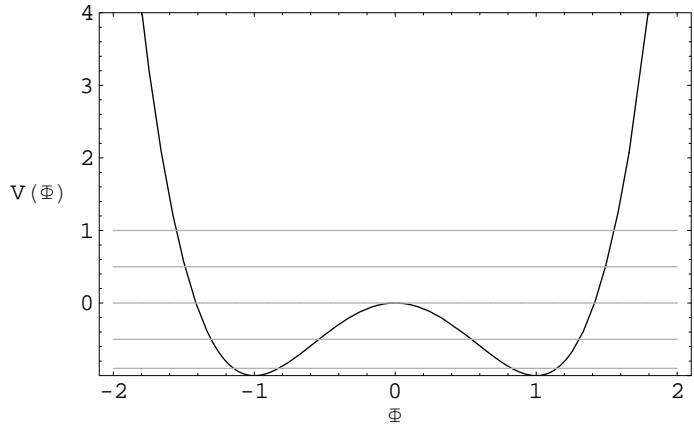}
   \end{center}
   \caption{The Higgs \emph{vev} potential~$V(\Phi)=\Phi^{4}- 2\Phi^{2}$.
   Straight horizontal lines correspond to $c=-0.9,-0.5,0.5,0,1$, with $c$ proportional
     to the Higgs vacuum total energy.}
   \label{fig:vacuum_pot}
\end{figure}

The starting point of our work is the assumption that the Higgs field is a
\emph{time-dependent} ``classical'' homogeneous condensate, $v = v(t)$.
Working in the unitary gauge ($\xi^a=0$), the Lagrangian \eqref{lagrH} becomes
\begin{eqnarray}
\label{lagr_phi,h}
{\mathcal L}_{H} 
=\frac{1}{2}\dot{v}^{2}
+\frac{M_{H}^{2}}{4} v^{2}
- \frac{\lambda}{8} v^{4}
+{\mathcal L}_{v,h}
,
\end{eqnarray}
where a dot denotes the derivative with respect to $t$ and
${\mathcal L}_{v,h}$ is a polynomial in $h$ and its derivatives.
We then introduce a dimensionless real scalar field $\Phi$ such that
\begin{equation}
v = 2\,\frac{M}{g}\,\Phi
\ ,
\end{equation}
and
\begin{equation}
H=\frac{1}{\sqrt{2}}\left[\frac{2M}{g}\Phi(t)+h(x)\right]\binom{0}{1}
.
\label{un_g}
\end{equation}
in order to  study the fluctuations around the constant value  $\Phi^2 = 1$, which is
at the basis of the SSB mechanism.
On neglecting all terms involving $h$ (which will be analysed in Section~\ref{back}),
the Euler-Lagrange equation for $\Phi$
becomes
\begin{equation}
\label{phi_eq_with_br}
 \ddot{\Phi} - \frac{M_{H}^{2}}{2}\, \Phi + \frac{M_{H}^{2}}{2}\,\Phi^{3}=0
 \ ,
\end{equation}
which can be put in dimensionless form by
rescaling the time as $\tau\equiv M_H t/2$ (primes will denote
derivatives with respect to $\tau$),
\begin{equation} 
\label{phi_eq}
\Phi'' - 2\Phi(1-\Phi^{2})=0.
\end{equation} 
After a simple integration we obtain
\begin{equation}
\label{phi_eq_1}
  \Phi'^{2} - 2\Phi^{2} + \Phi^{4}  \equiv \Phi'^{2}+V(\Phi)=c,
\end{equation}
where the integration constant
$c\in\mathbb{R}$ is a function of the initial conditions.
From a physical point of view, $c$ is proportional to the total vacuum
energy\footnote{It is easy to see that the integral over space
of the Hamiltonian density obtained from \eqref{lagr_phi,h} gives
the total Higgs vacuum energy $E_{\Phi}={M_{H}}c/{\lambda}$.
We shall thus refer to $c$ as the ``Higgs vacuum total energy'' for
brevity.}
and $V(\Phi)$ is the potential in which this vacuum is ``moving'' (see fig.~\ref{fig:vacuum_pot}).
\par
The equation \eqref{phi_eq_1} can be formally solved by separating variables,
which yields
\begin{equation} \label{integral}
  \tau - \tau_{0} = \pm \int_{\Phi_{0}}^{\Phi} \frac{d\Phi}{\sqrt{-\Phi^{4}
  +2\Phi^{2} + c}} \equiv \int_{\Phi_{0}}^{\Phi} f(\Phi)\,d\Phi\, .
\end{equation}
Since we are interested in real solutions of Eq.~\eqref{phi_eq_1}, we
must restrict to the case
\begin{equation}
\label{real_dom}
 \left( -\Phi^{4} + 2\Phi^{2} + c \right)>0,
\end{equation}
which corresponds to positive kinetic energy for $\Phi$ (see eq.~\eqref{phi_eq_1}).
For any $c$, the stationary points of $f(\Phi)$ are given by $ \Phi=0, \: \pm 1$,
for which the potential $V(\Phi)$ has a maximum and two minima rispectively. 
Moreover, note that for $c=0$ and $c=-1$ the integral in Eq.~\eqref{integral} is
not defined at these points.
We shall therefore consider the following cases:
\begin{enumerate}
\item
$c \leq -1$.
Since $c=-1$ corresponds to the absolute minimum in the vacuum total energy,
this case is not physically relevant;
\item
\label{-1c0}
$-1<c<0$.
The system exhibits closed trajectories in phase space and a periodic evolution;
\item
$c=0$.
The motion is along the separatrix and the system leaves the equilibrium point
$\Phi=0$ with an exponentially growing velocity;
\item
$c>0$.
The system again exhibits closed trajectories in phase space but with periods longer
than those in case~\ref{-1c0}.
\end{enumerate}
It can further be shown that the solutions for all four cases can be connected by analytic
continuation\footnote{These solutions are well known in the literature for the $\phi^{4}$ theory
in $(1+1)$ and $(3+1)$-dimensions~\cite{phi4}.}.
Therefore, in what follows, we shall use the simplest form of such solutions
\cite{Gradsh_Ryz} (which
corresponds to $c>0$) for the sake of simplicity, namely
\begin{equation}
\label{phi_t}
 \Phi(\tau)=L_{m}\, \cn \,(R_{m}\tau,m)
,
\end{equation}
where $L_{m}\equiv\sqrt{\frac{2m}{2m-1}}$ and
$R_{m}\equiv\sqrt{\frac{2}{2m-1}}$.
The parameter $m$ is given by
\begin{equation}
\label{ellip-parameter}
  m=\frac{1+\sqrt{1+c}}
         {2\sqrt{1+c}} ,
\end{equation}
and is a function of the initial conditions because so is the
vacuum energy $c$.
The function $\cn(z,m)$ is called Jacobi Elliptic function and 
has the following properties \cite{table_function}:
\begin{eqnarray} \label{prop_cn}
&& \textrm{periods:} \qquad 4\textbf{K}, \quad 2\textbf{K}+2i\textbf{K'} \\
&& \textrm{zeros:}    \qquad (2l+1)\textbf{K}+2ni\textbf{K'} \\
&& \textrm{poles:}    \qquad  \beta_{l,n}=2l\textbf{K}+(2n+1)i\textbf{K'}\, ,
\end{eqnarray}
where $l$, $n$ are integers and $\textbf{K}(m)$ is a special case of
the complete elliptic integral of the first kind $F(\theta,m)$,
\begin{equation}
\textbf{K}(m)=F\left( \frac{\pi}{2},m\right) \qquad \textbf{K}'(m)
=F\left(\frac{\pi}{2},m'\right) ,
\end{equation}
with $m'=1-m$.
From \eqref{prop_cn} we deduce that the period along the real (imaginary)
axis of the function in Eq.~\eqref{phi_t} is given by
\begin{equation}
\label{phi-periods}
 T=\left\{
 \begin{array}{ll}
 2\sqrt{2}\sqrt{2m-1}F\left( \frac{\pi}{2},m\right),
             &\textrm{for}\; m<1
             \quad (m>1)
             \\
   &\\
 \sqrt{2}\sqrt{2m-1}\left[F\left( \frac{\pi}{2},m\right)+
                       iF\left( \frac{\pi}{2},1-m\right)\right],
             &\textrm{for}\; m>1
             \quad (m<1)\,.
\end{array}\right.
\end{equation}
It is also important to note that the solution \eqref{phi_t} describes the evolution
in time of the Higgs vacuum $\Phi$ for every initial conditions
$(\Phi_{0},\Phi'_{0})$ only after a suitable time \emph{shift} (so that $\Phi'(\tau=0)=0$).
%
%
\section{Fermions in a time dependent Higgs \emph{vev}}
\label{sec-fermion_theory}
\setcounter{equation}{0}
In the Standard Model the coupling of a generic fermion field $\psi$ to the
Higgs scalar field $H$ and gauge fields is described by the Lagrangian density
terms
\cite{SMpassarino}
\begin{eqnarray}
\label{lagr_ferm_1}
\mathcal{L}_{f}&\!\!=\!\!&
i\bar{\psi}_{L}\gamma^{\mu}
      \left(\partial_{\mu}-ig\frac{\vec{\sigma}}{2}\cdot \vec{A}_{\mu}(x)
             -ig'\frac{Y_{W}}{2}B_{\mu}\right)\psi_{L}
 \nonumber
 \\
&&
 +i\bar{\psi}_{R}\gamma^{\mu}\left(\partial_{\mu}-ig'\frac{Y_{W}}{2}B_{\mu}
                                        \right)\psi_{R}
           -G\left(\bar{\psi}_{L}H\psi_{R}+\bar{\psi}_{R}H^{\dagger}\psi_{L}\right)
           ,
\end{eqnarray}
where $\psi_{L}$ is a \emph{left handed} isospin doublet,
$\psi_{R}$ a \emph{right handed} isospin singlet and $G$ the coupling constant between
the Higgs boson and the fermions.
Neglecting the gauge fields, in the unitary gauge (\ref{un_g}),
the previous Lagrangian becomes
\begin{equation}
\label{lagr_fermioni}
 \mathcal{L}_{f}= i\bar{\psi}\gamma^{\mu}\partial_{\mu}\psi
     -\frac{G}{\sqrt{2}}\frac{2M}{g}\Phi(t)\bar{\psi}\psi-\frac{G}{\sqrt{2}}h(x)\bar{\psi}\psi
,
\end{equation}
with $\psi=\psi_{L}+\psi_{R}$.
Introducing the ``mass parameter''\footnote{Since there are no stationary states in a
time-dependent external field $\Phi(t)$, the mass is strictly speaking ill-defined.
We shall however refer to the function $m_{f}(t)$ as a time dependent mass.}
\begin{equation}
\label{m(t)}
 m_{f}(t)\equiv \frac{G}{\sqrt{2}}\frac{2M}{g}\Phi(t)=
       \frac{G}{\sqrt{2}}\frac{M_{H}}{\sqrt{\lambda}}\Phi(t)\equiv m_{f}\Phi(t)
,
\end{equation}
with $\Phi(t)$ given by Eq.~\eqref{phi_t}, the equation of motion for $\psi$ is given by
\begin{equation}
\label{eq_Dirac_1}
\left[i\gamma^{\mu}\partial_{\mu} - m_{f}(t) -\frac{G}{\sqrt{2}}\,h(x)\right] \psi(x)=0
.
\end{equation}
For our purposes, the last term above can be neglected with respect to the second
term proportional to the Higgs condensate and one finally obtains 
\begin{equation}
\label{eq_Dirac}
 \left[i\gamma^{\mu}\partial_{\mu}-m_{f}(t)\right]\psi(\vec{x},t)=0
\end{equation}
which is a Dirac equation with a \emph{time dependent mass}.
In a similar manner, one finds that the equation of motion for
$\bar{\psi}=\psi^{\dagger}\gamma^{0}$ is the Hermitian conjugate of
\eqref{eq_Dirac}.
\par
The spinor field is normalized in such a way that
$\int d^{3}x\, \psi^{\dagger}(\vec{x},t)\psi(\vec{x},t)=1$ and, in the
Heisenberg picture, it becomes a field operator
with the usual anticommutation rules
$\{\psi(\vec{x},t),\psi^{\dagger}(\vec{x}\,',t)\}=\delta(\vec{x}-\vec{x}\,')$.
We can then expand $\psi$ as
\begin{equation}
\label{sviluppo_psi}
 \psi(x)=\int \frac{d^{3}k}{(2\pi)^{3/2}}\sum_{s=\pm}
          e^{i\vec{k}\cdot \vec{x}}
             \left[U_{s}(\vec k,t)a_{s}(\vec k)+
                  V_{s}(-\vec k,t)b_{s}^{\dagger}(-\vec k)\right]
,
\end{equation}
where $s=\pm$ is the helicity,
\begin{equation}
\label{a_b_comm_rules}
 \{a_{s}(\vec k),a^{\dagger}_{s'}(\vec k')\}=\{b_{s}(\vec k),b^{\dagger}_{s'}(\vec k')\}=
   \delta_{ss'}\delta(\vec{k}-\vec{k}')
\end{equation}
and
\begin{equation}
\label{norm_UV}
\begin{array}{l}
U_{s}^{\dagger}(\vec k,t)U_{s'}(\vec k,t)=
V_{s}^{\dagger}(\vec k,t)V_{s'}(\vec k,t)=\delta_{ss'}
\\
\\
U_{s}^{\dagger}(\vec k,t)V_{s'}(\vec k,t)=V_{s}^{\dagger}(\vec k,t)U_{s'}(\vec k,t)=0
\ ,
\end{array}
\end{equation}
The vacuum state $|0\rangle$ is as usual defined by the relations
\begin{equation}
\label{vacuum_state}
 a_{s}(\vec k)|0\rangle=b_{s}(\vec k)|0\rangle=0
 .
\end{equation}
With no time dependence in the theory ({i.e.}, for $m_{f}(t)=m_{f}$ constant),
$U_{s}(\vec k)$ and $V_{s}(\vec k)$ would be eigenstates of the operator
$\vec{\gamma}\cdot \vec{k}$ with eigenvalues $m_{f}$ and $-m_{f}$
respectively.
The spinors in momentum space $U_{s}(\vec k,t)$ and $V_{s}(\vec k,t)$ satisfy the
charge coniugation relation
\begin{equation}
\label{coniug_carica}
 \mathscr{C}\bar{U}_{s}^{T}(\vec k,t)=V_{s}(-\vec k,t)
 ,
\end{equation}
with $\mathscr{C}=i\gamma_{0}\gamma_{2}$\footnote{We are using the gamma
matrices 
\begin{equation*}
\gamma^{0}=\left(\begin{array}{c|c} \mathbb{I}&0\\
                                      \hline
                                    0&\mathbb{-I}
                  \end{array}\right)
                  ,
\qquad
\gamma^{j}=\left(\begin{array}{c|c}  0 &-\sigma^{j}\\
                                       \hline
                                    \sigma^{j}&0
                 \end{array}\right),
\end{equation*}
where $\sigma^{j}$, $j=1,2,3$ are Pauli matrices and $\mathbb{I}$ is the
$2\times2$ identity matrix.}.
\par
It is now convenient to introduce two new scalars defined by
\begin{equation}
\label{uv_xpxm}
\begin{array}{l}
 U_{s}(\vec k,t)=\left[i\gamma^{0}\partial_{0}+\vec{\gamma} \cdot \vec{k}+m_{f}(t)\right]
                     X_{k}^{\p}(t)u_{s}
 \\
 \\
 V_{s}(\vec k,t)=\left[i\gamma^{0}\partial_{0}-\vec{\gamma} \cdot \vec{k}+m_{f}(t)\right]
                     X_{k}^{\m}(t)v_{s}
 ,
\end{array}
\end{equation}
where
\begin{equation}
 u_{s}=\binom{\chi_{s}}{0} \qquad v_{s}=\binom{0}{\eta_{s}}\,,
\end{equation}
with $\chi_{s}^{\dagger}\chi_{s}=1$ and $\eta_{s}=-i\sigma_{2}\chi_{-s}$,
are eigenvectors of $\gamma^{0}$ with
eigenvalues +1 and -1, respectively.
Note that $\mathscr{C}\bar{u}^{T}_{s}=v_{-s}$ and
Eq.~\eqref{coniug_carica} is thus identically satisfied.
With these notations, Eq.~\eqref{eq_Dirac} yields
\begin{equation}
\label{eq_modi}
  \ddot{X}_{k}^{\tpm}(t)+\left[\Omega_{k}^{2}(t)\mp i\dot{m}_{f}(t)\right]
     X_{k}^{\tpm}(t)=0
     ,
\end{equation}
which is of the harmonic oscillator type with the complex and
(doubly-)periodic frequency
\begin{equation}
\Omega_{k}^{2}(t)\mp i\dot{m}_{f}(t)\equiv k^{2}+m_{f}^{2}(t)\mp i\dot{m}_{f}(t)
.
\end{equation}
\par
Let us now assume that the Higgs \emph{vev} remains constant and equal
to $\Phi(0)$ for $t\leq 0$.
Consequently, the mass $m_{f}(t)$ will also be constant at negative times and
one just has plane waves for $t\leq 0$. 
The evolution for $t>0$ is then obtained by imposing the
following initial conditions at $t=0$\footnote{We have set the
momentum $\vec{k}=(0,0,k)$.}
\begin{equation}
\label{condizioni_iniziali}
\left\{
\begin{array}{l}
          X_{k}^{\tpm}(0)=\left\{2\Omega_{k}(0)\left[\Omega_{k}(0)+m_{f}(0)\right]\right\}^{-1/2}
          \\
          \\
        \dot{X_{k}}^{\tpm}(0)=\mp i \, \Omega_{k}(0)X_{k}^{\tpm}(0)
        .
\end{array}
\right.
\end{equation} 
These together with Eq.~\eqref{eq_modi} give
\begin{equation}
\label{rel_xp_xm}
 X_{k}^{\m}(t)=\left(X_{k}^{\p}(t)\right) ^{*}
 ,
\end{equation}
so that positive and negative energy modes are not independent and
we shall then consider mostly the equation for $X_{k}^{\p}$ for simplicity.
\par
It can be showed that if $f_{1}(t)$ and $f_{2}(t)$ are two
arbitrary solutions of Eq.~\eqref{eq_modi} with the sign $\p$ (or,
equivalently, with $\m$), the quantity
\begin{equation}
\label{invariante}
 I[f_{1},f_{2}]\equiv \Omega_{k}^{2}(t)f_{1}^{*}f_{2}+\dot{f_{1}^{*}}\dot{f_{2}}
         +im_{f}(t)\left(f_{1}^{*}\dot{f_{2}}-\dot{f_{1}^{*}}f_{2}\right)
\end{equation} 
is a constant of motion and one can then prove the stability of
any arbitrary solutions \cite{mostepanenko_frolov}.
Finally, note that if $f_{1}(t)=f_{2}(t)=X_{k}^{\p}(t)$ the relation \eqref{invariante}
takes the form
\begin{equation} \label{invariante=1}
\left|\dot{X}_{k}^{\p}\right|^{2}
+\Omega_{k}^{2}\left|X_{k}^{\p}\right|^{2}
   +im_{f}(t)\left(X_{k}^{\p *}\dot{X}_{k}^{\p} - \dot{X}_{k}^{\p *}X_{k}^{\p}\right)
   =1
   ,
\end{equation}
which is also a consequence of the fact that
$U_{s}(\vec k,t)$ and $V_{s}(\vec k,t)$ are evolved by the Hermitian operators
$i\gamma^{0}\partial_{0}\mp \vec{\gamma} \cdot \vec{k}-m_{f}(t)$.
%
\subsection{Fermion solutions and physical quantities}
%
The Hamiltonian operator for a fermion field can in general be written as
\begin{equation}
\label{def_ham}
 \mathcal{H}(t)
  =i\int d^{3}x\, \psi^{\dagger}(\vec{x},t)\dot{\psi}(\vec{x},t)
.
\end{equation}
Inserting the expansion \eqref{sviluppo_psi} and using
Eq.~\eqref{norm_UV}, this becomes 
\begin{eqnarray}
 \mathcal{H}(t)&\!\!=\!\!&\int d^{3}k \sum_{s}\left\{
   \left[iU_{s}^{\dagger}(\vec k,t)\dot{U}_{s}(\vec k,t)\right]a_{s}^{\dagger}(\vec k)a_{s}(\vec k)
   + \left[iV_{s}^{\dagger}(-\vec k,t)\dot{V}_{s}(-\vec k,t)\right]b_{s}(-\vec k)b_{s}^{\dagger}(-\vec k)
   \right. 
 \nonumber
 \\
 &&
   \left.
   +\left[iV_{s}^{\dagger}(-\vec k,t)\dot{U}_{s}(\vec k,t)\right]b_{s}(-\vec k)a_{s}(\vec k)
   +\left[iU_{s}^{\dagger}(\vec k,t)\dot{V}_{s}(-\vec k,t)\right]a_{s}^{\dagger}(\vec k)b_{s}^{\dagger}(-\vec k)
\right\}
,                                   
\end{eqnarray}
where we have integrated on $\vec{x}$ and one of the momenta.
Taking into account Eq.~\eqref{uv_xpxm}, we end up with
\begin{eqnarray}
\label{ham_non_diag}
 \mathcal{H}(t)&\!\!=\!\!&
 \int d^{3}k\sum_{s=\pm}\Omega_{k}(t)\left\{
     E(k,t)\left[a_{s}^{\dagger}(\vec k)a_{s}(\vec k)-b_{s}(-\vec k)b_{s}^{\dagger}(-\vec k)\right]
  \right.
  \nonumber
  \\
  &&
  \qquad
  \left.
     +F(k,t) b_{s}(-\vec k)a_{s}(\vec k)+
      F^{*}(k,t)a_{s}^{\dagger}(\vec k)b_{s}^{\dagger}(-\vec k)\right\}
,
\end{eqnarray}
where
\begin{equation}
\label{E_F}
\begin{array}{l}
E(k,t)= \frac{2k^{2}}{\Omega_{k}(t)}
                 \im\left[X_{k}^{\p}(t)\dot{X}_{k}^{\p *}(t)\right]
                 +\frac{m_{f}(t)}{\Omega_{k}(t)}
                 \ ,
                  \\
                  \\
F(k,t)\equiv  \frac{k}{\Omega_{k}(t)}\left[(\dot{X}_{k}^{\p}(t))^{2}+
                            \Omega_{k}^{2}(t)(X_{k}^{\p }(t))^{2}\right]
\end{array}
\end{equation}
and\footnote{This is a consequence of \eqref{invariante=1}.} 
\begin{equation}
\label{E+F=1}
E^{2}(k,t)+|F(k,t)|^{2}=1
.
\end{equation}
From Eq.~\eqref{E_F}, using \eqref{condizioni_iniziali}, it is
possible to see that $E(k,0)=1$ and $F(k,0)=0$, therefore $\mathcal{H}(t=0)$ is
diagonal.
In fact, we have assumed that for $t\leq 0$ the vacuum is constant and there is no explicit time
dependence in the theory.
\par
The Hamiltonian \eqref{ham_non_diag} can be diagonalized
\emph{at every time} using a canonical Bogoliubov trasformation \cite{Bogoliubov}.
As a matter of fact, the necessary conditions for this kind of diagonalization are
ensured by the relations \cite{Mostepanenko_book}
\begin{equation}
\begin{array}{l} 
iV_{s}^{\dagger}(-\vec k,t)\dot{V}_{s}(-\vec k,t)
= iU_{s}^{\dagger}(\vec k,t)\dot{U}_{s}(\vec k,t)
= \Omega_{k}(t)E(k,t)
\\
\\
 \left[iU_{s}^{\dagger}(\vec k,t)\dot{V}_{s}(-\vec k,t)\right]^{*}
 = iV_{s}^{\dagger}(-\vec k,t)\dot{U}_{s}(\vec k,t)
 =\Omega_{k}(t)F(k,t)
 . 
\end{array}
\end{equation}
We now introduce \emph{time-dependent} creation and annihilation operators,
\begin{equation}
\label{bog^-1}
 \left( \begin{array}{c}
         \tilde{a}_{s}(\vec k,t)\\ 
         \tilde{b}_{s}^{\dagger}(\vec k,t)
        \end{array}
 \right)=
 \left[
   \begin{array}{cc}
          \alpha(k,t) & \, \beta(k,t)\\
         -\beta^{*}(k,t)  & \, \alpha^{*}(k,t)
   \end{array}
 \right]
        \left(\begin{array}{c}
                a_{s}(\vec k)\\ 
                b^{\dagger}_{s}(-\vec k)
        \end{array}
        \right)\equiv
  \mathcal{A}(k,t)
         \left( \begin{array}{c}
                a_{s}(\vec k)\\ 
                b^{\dagger}_{s}(-\vec k)
                 \end{array}
         \right)
         ,
\end{equation}
and the condition that this be a non singular canonical trasformation
requires that $\mathcal{A}$ is a special unitary matrix,
\begin{equation}
\label{alpha+beta=1}
\left|\alpha(k,t)\right|^{2}+\left|\beta(k,t)\right|^{2}=1
.
\end{equation}
We have thus shown that $SU(2)$ is the
dynamical symmetry group for fermion creation in a homogeneus non stationary
scalar field and the time-dependent vacuum $|0\rangle_{t}$ (see below) is
a generalized coherent state built on this group \cite{perelomov}.
\par 
The hamiltonian also takes the diagonal form
\begin{equation}
\label{ham_diag}
  \mathcal{H}(t)=\int d^{3}k\sum_{s=\pm}\Omega_{k}(t)\left[
     \tilde{a}_{s}^{\dagger}(\vec k,t)\tilde{a}_{s}(\vec k,t)
         -\tilde{b}_{s}(-\vec k,t)\tilde{b}_{s}^{\dagger}(-\vec k,t)
                                                    \right]
                                                    \ , 
\end{equation}
if the coefficients of the canonical trasformation are such that
\begin{equation}
\label{alpha,beta_E,F}
\begin{array}{l}
\strut\displaystyle
\left|\beta(k,t)\right|^{2}=\frac{1-E(k,t)}{2}
\\
\\
\strut\displaystyle
 \frac{\alpha(k,t)}{\beta(k,t)}=\frac{F(k,t)}{1-E(k,t)}=\frac{1+E(k,t)}{F^{*}(k,t)}
 ,
\end{array}
\end{equation}
which are indeed compatible with the condition \eqref{alpha+beta=1}
thanks to \eqref{E+F=1}.
\par
It is now possible to use the operators $\tilde{a}$ e $\tilde{b}$ to
define time dependent Fock spaces, each of them built from the zero
(quasi)particle state at the time $t$,
$\tilde{a}_{s}(\vec k,t)|0_{t}\rangle=0=\tilde{b}_{s}(\vec k,t)|0_{t}\rangle$,
which, at $t=0$, are equal to $a_{s}(\vec k)|0\rangle=0=b_{s}(\vec k)|0\rangle$.
These relations mean that a quantized fermion field interacting with a
classical external field $\Phi(t)$ can be represented
at every time as a free field, with a corresponding redefinition af the
particle concept and vacuum state.
Moreover, one can show that diagonalizing the Hamiltonian
\eqref{ham_non_diag} is equivalent to finding exact solutions of the Heisenberg
equations of motion and all the matrix elements 
(expectation values of physical observables) of interest can be written
in terms of the coefficients of the Bogoliubov trasformation~\eqref{bog^-1}
\cite{Mostepanenko_book}.
\par
For example, the vacuum expectation value of the
(quasi-)particle number operator is given by
\begin{equation}
  N_{k}(t) \equiv \label{N_{k}(t)}
  \langle 0|\tilde{a}_{s}^{\dagger}(\vec k,t) \tilde{a}_{s}(\vec k,t)|0\rangle 
   =\left|\beta(k,t)\right|^{2}\langle0|b_{s}(\vec k)b_{s}^{\dagger}(\vec k)|0\rangle
  =\left|\beta(k,t)\right|^{2}\delta(\vec 0),
\end{equation}
where we have used Eqs.~\eqref{bog^-1} and \eqref{vacuum_state}.
From the previous relation we see that the number of created (quasi)particle pairs is
spin-independent because the homogeneus field $\Phi$ is isotropic \cite{popov_marinov}.
If we put the system in a finite volume $V$, we must replace
$\delta(\vec 0)$ in \eqref{N_{k}(t)} with $\delta_{\vec{k}\vec{k}}=1$.
The (quasi)particle density at the time $t$ is thus
given by \footnote{Of course, the same result holds for antifermions.}
\begin{eqnarray}
\label{def_NO}
 n(t)=
  \langle0|\frac{1}{V}\sum_{s=\pm}\int \frac{d^{3}k}{(2\pi)^{3}} N_{k}(t)|0\rangle
      =\frac{2}{(2\pi)^{3}}\int d^{3}k\langle0|N_{k}(t)|0\rangle 
     =\frac{1}{\pi^{2}}\int dk\,k^{2}|\beta(k,t)|^{2}
    ,
\end{eqnarray}
which is different from zero whenever the Hamiltonian is not diagonal in terms
of the operators $a$ and $b$.
The occupation number of fermions created with a given momentum $\vec k$ will be
$n_{k}(t)=|\beta(k,t)|^{2}$, and the condition \eqref{alpha+beta=1} ensures
that the Pauli principle is respected at every time \cite{Mostepanenko_book,pel-sor}. 
\par
In order to implement numerical methods, it is useful to cast some of the
previous expressions in dimensionless form.
We thus introduce the following quantities:
\begin{equation}
\label{parametri_adim}
    \tau \equiv \frac{M_{H}}{2}\,t \,, 
         \qquad \quad \quad
        \kappa \equiv \frac{2\,k}{M_{H}}\,,
         \quad \quad \quad 
   q \equiv 2\,\frac{G^{2}}{\lambda}=4\frac{m_{f}^{2}}{M_{H}^{2}},
\end{equation}
where $m_{f} \equiv m_{f}(\tau\rightarrow -\infty)= m_{f}(0)$.
On further multiplying by ${4}/{M_{H}^{2}}$, Eq.~\eqref{eq_modi}
takes the dimensionless form
\begin{equation}
\label{eq_modi_adim}
  X_{k}^{\tpm''}(\tau)+\left[\kappa^{2}+q\,\Phi^{2}(\tau,m)
    \mp i\sqrt{q}\,\Phi'(\tau,m)\right] X_{k}^{\tpm}(\tau)=0 .
\end{equation}
Moreover, if we define the dimensionless frequency
\begin{equation}
\omega_{\kappa}
\equiv
\frac{2\,\Omega_{k}}{M_{H}}
=\sqrt{\kappa^2+q \,\Phi^2}
\ ,
\label{om_f}
\end{equation}
the
initial conditions \eqref{condizioni_iniziali} become
\begin{equation}
\label{condizioni_iniziali_adim}
\left\{
\begin{array}{l}
   X_{\kappa}^{\tpm}(0)=
     \left\{2\,\omega_{\kappa}(0)[\omega_{\kappa}(0)+\sqrt{q}\,\Phi(0,m)]\right\}^{-1/2}
 \\
 \\
   X_{\kappa}^{\tpm'}(0)=\mp i \, \omega_{\kappa}(0)
                 X_{\kappa}^{\tpm}(0)
   . 
\end{array}
\right.
\end{equation}
Note that, despite this is not explicitly indicated, the frequency and the
initial conditions are also functions of the amplitude $m$ of the elliptic function,
that is of the vacuum energy $c$ (see Eq.~\eqref{ellip-parameter}).
On using  \eqref{alpha,beta_E,F}, \eqref{E_F} and  \eqref{parametri_adim} we
finally obtain
\begin{equation}
\label{numero_occupazione_adim}
n_{\kappa}(\tau)=\frac{1}{2}
     -\frac{\kappa^{2}}{\omega_{\kappa}(\tau)} 
      \im \left[X_{\kappa}
         (X'_{\kappa})^{*}\right]
        -\frac{\sqrt{q}\,\Phi(\tau,m)}{2 \omega_{\kappa}(\tau)}
        ,
\end{equation}
which gives the occupation number for every mode $\kappa$ as a function of the
solutions $X_{\kappa} \equiv X_{\kappa}^{\p}$~\footnote{This definition is not restrictive
since we only need either $X_{\kappa}^{\p}$ or $X_{\kappa}^{\m}$ 
to calculate $n_{\kappa}(\tau)$.} of Eq.~\eqref{eq_modi_adim}.
Analogously, the (dimensionless) energy density will be
\begin{equation}
\label{densita_energia_adim}
  \tilde{\rho}_{\psi}(\tau)=\frac{1}{2\,\pi^{2}}\int d\kappa\,\kappa^{2}
        \omega_{\kappa}(\tau) \, n_{\kappa}(\tau)
        .
\end{equation}
Note that $n_{\kappa}(0)=\tilde{\rho}_{\psi}(0)=0$ thanks to the initial
conditions \eqref{condizioni_iniziali_adim}.
\par
\par
A very useful result follows from the periodicity of the
vacuum~\footnote{We clearly refer to the period of $\Phi$ along the real axis
(see Eq.~\eqref{phi-periods}).},
$\Phi(\tau)=\Phi(\tau+T)$, which remarkably simplifies the evaluation of the occupation
number and shows, although in an approximate way, its explicit
time dependence \cite{mostepanenko_frolov}.
If we define
\begin{equation}
 \omega_{\kappa} \equiv \omega_{\kappa}(\tau\rightarrow -\infty)=\omega_{\kappa}(0)
,
\end{equation}
an approximate expression for $n_{\kappa}(\tau)$ is given by~\footnote{We shall see in
Section~\ref{fermi_pro} that Eq.~\eqref{inviluppo_n(t)_adim} is actually exact
at $\tau=n\,T$ for any positive integer $n$.}
\begin{equation}
\label{inviluppo_n(t)_adim}
   \hat{n}_{\kappa}(\tau)=\frac{\kappa^{2}}{\omega_{\kappa}^{2}}
       \frac{\{\im[X_{\kappa}^{(1)}(T)]\}^{2}}
            {\sin^{2}(d_{\kappa})}
       \sin^{2}\left(d_{\kappa}\frac{\tau}{T}\right)
    \equiv
      F_{\kappa} \sin^{2}\left(\nu_{\kappa}\tau\right)
  ,
\end{equation}
%
where $X_{\kappa}^{(1)}(\tau)$ satisfies Eq.~\eqref{eq_modi_adim}
with initial conditions $X_{\kappa}^{(1)}(0)=1$, $X_{\kappa}^{(1)'}(0)=0$
and $d_{\kappa}$ is such that $\cos (d_{\kappa})=\re[X_{\kappa}^{(1)}(T)]$. 
According to Eq.~\eqref{inviluppo_n(t)_adim} the number density of fermions
produced depends periodically on $\tau$ for all $\kappa$.
So, on average, this density does not depend on the time during which the external
field is turned on.
The physical meaning of this result was pointed out by V.~S.~Popov through a
quantum mechanics analogy \cite{mostepanenko_frolov}.
Finally observe that in our case the time scale is fixed by the factor
${2}/{M_{H}}$ and if $M_{H}\sim 10^{2}\,$ GeV then~\footnote{We recall that
$1\,$GeV$^{-1} \sim 6.582 \cdot 10^{-25}\,$s for $\hbar=c=1$.}
${2}/{M_{H}}\sim 1.3\cdot 10^{-26}\,$s.
%
%
\subsection{Band structure}
%
A non adiabatic quantum effect arises from the explicit time dependence of the
frequency in Eq.~\eqref{eq_modi_adim}, and this leads to the production of
particles.
When the time dependence is periodic, one usually speaks of
\emph{parametric resonance}.
It is then clear that the quantity $q$ has the
role of a \emph{resonance parameter} due to the fact that the time-dependent terms
in Eq.~\eqref{eq_modi_adim} are proportional to $q$.
\par
\begin{figure}[t]
   \centering
   \includegraphics[scale=1.1]{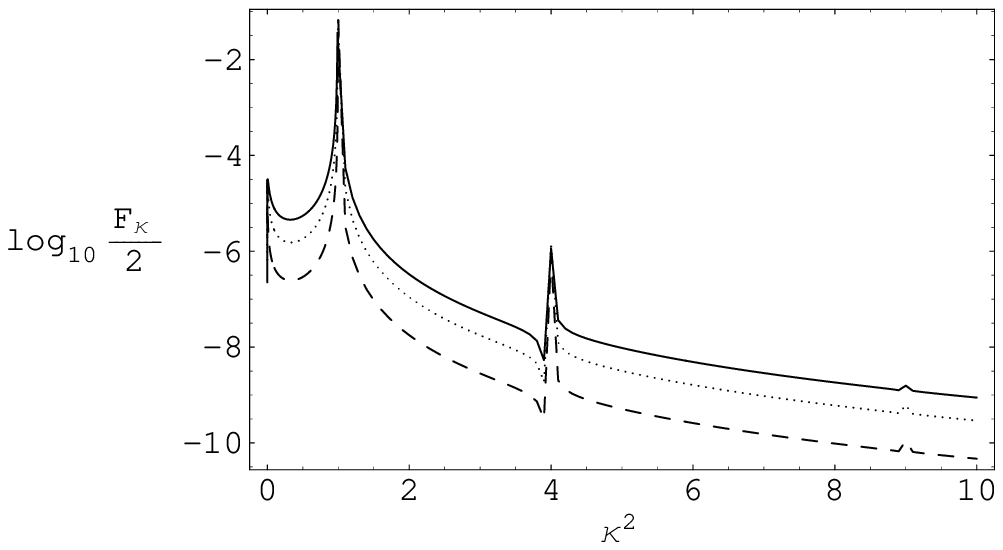} 
 \caption{Mean occupation number ${F_{\kappa}}/{2}$ as a function of
   $\kappa^{2}$ for the \emph{bottom} and $M_{H}=115\,$GeV (solid line),
   $200\,$GeV (dotted line), $500\,$GeV (dashed line).
   Note that the peaks get narrower with increasing $\kappa^{2}$.
   Similar plots are obtained by varying
   the resonance parameter $q$ at $\kappa^{2}$ fixed as
for decreasing $q$ the bandwidth shrinks.}
\label{picchi(M_H)} 
\end{figure}

We are interested in values of $q$ and $\kappa$ which give solutions of
the mode equation associated to particle production, identified by a mean occupation number
(see Eq.~\eqref{inviluppo_n(t)_adim})
\begin{equation}
\label{no_medio}
 \bar{n}_{\kappa} \equiv \langle n_{\kappa}\rangle_{\tau}
                  = \frac{F_{\kappa}}{2}
\end{equation}
different from zero.
The result is shown in Fig.~\ref{cartainst} in which
every peak corresponds to $\bar{n}_{\kappa}=1/2$:
\begin{figure}[t]
   \centering
   \subfigure
   {\includegraphics[scale=0.80]{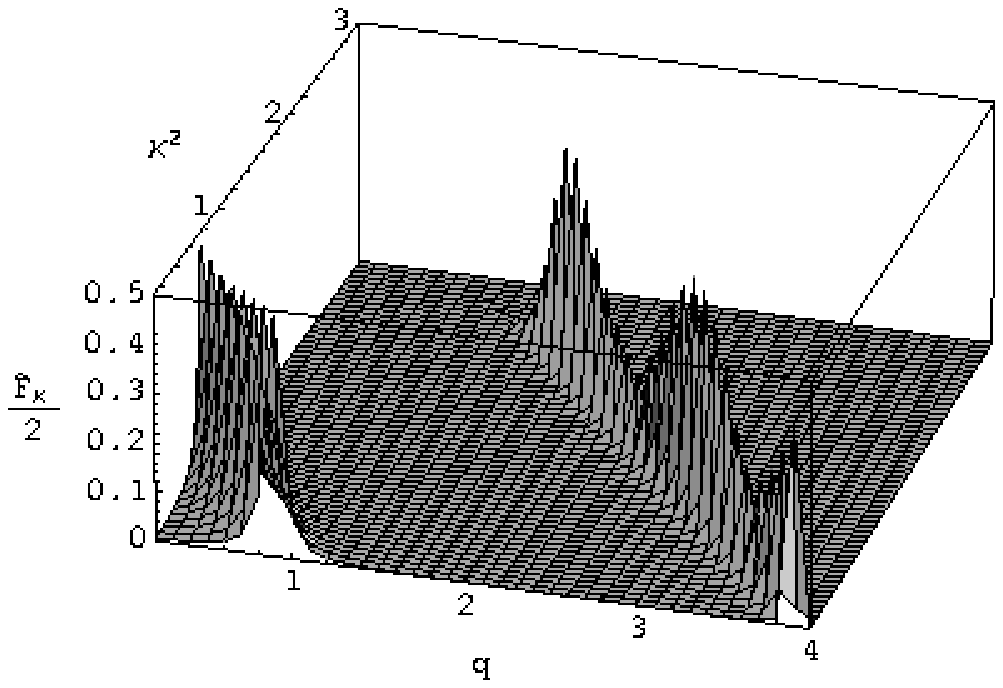}}
   \subfigure
   {\includegraphics[scale=0.80]{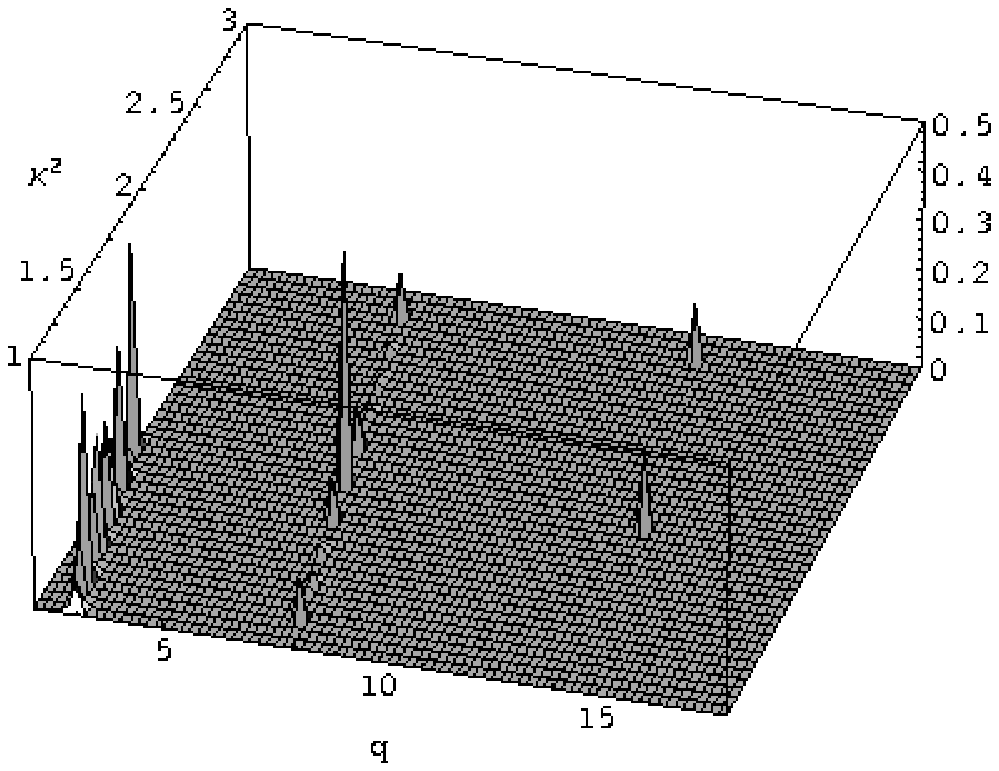}}\\
  \subfigure
  {\includegraphics[scale=0.75]{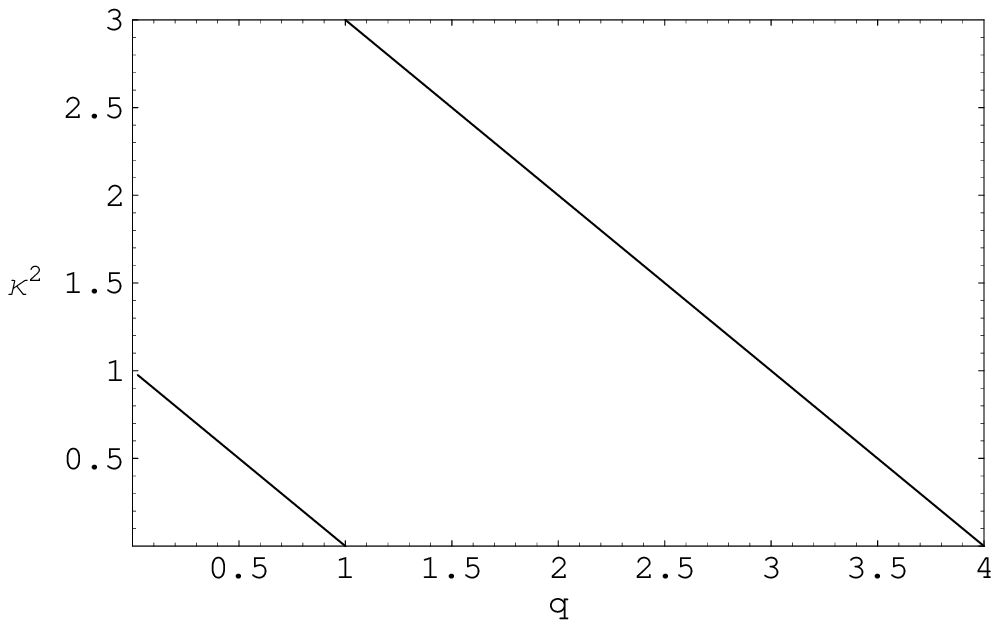}}
  \subfigure
  {\includegraphics[scale=0.75]{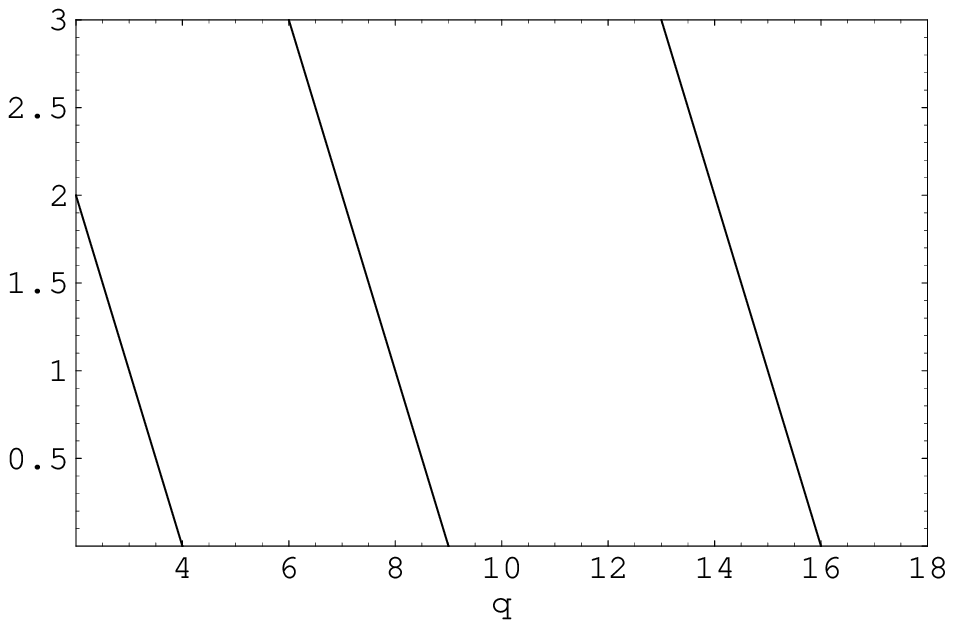}}
\caption{Production chart for fermions coupled to the Higgs.
Upper graphs show the occupation number as a function of the
parameters $q$ and $\kappa^2$.
Lower plots show in details the lines in the plane $(q,\kappa^{2})$ along
which the mean occupation number takes its maximum values.
Note that for $\kappa \simeq 0$ we have $q_{n}\simeq n^{2}$.}
\label{cartainst}
\end{figure}
note the band structure in the plane $(q,\kappa^{2})$.
The left plot in the upper part of Fig.~\ref{cartainst} shows the first and second bands
while the right plot displays bands from the second to the fourth.
Moving along a band, $\bar{n}_{\kappa}$ oscillates between $0$ and ${1}/{2}$.
Moreover, the bands get narrower with increasing $\kappa^{2}$ for a given value of $q$
and after several bands they shrink to a negligible width, as we show in
Fig.~\ref{picchi(M_H)} with a plot of the mean occupation number
as a function of $\kappa^{2}$ for the \emph{bottom} quark.
%
%
%
\par
To obtain the curves in the lower graphs of Fig.~\ref{cartainst} we have studied
the mean occupation number along straigh lines 
$\kappa^{2}=\vartheta \, q$, with $\vartheta$ constant.
First of all, for $\vartheta\approx 0$~\footnote{Note that if $\vartheta=0$, $\kappa^{2}=0$
and $\bar{n}_{\kappa}=0$ with no production.} the peaks are located on the $q$ axis
at $q_{n}(\vartheta\approx 0)=n^{2}$, with $n$ a positive integer.
We then made the ansatz
$q_{n}(\vartheta)=n^{2}\,\Gamma(\vartheta)$, with $\Gamma(0)=1$,
for the position of the peaks $q_{n}$ along the generic line $\kappa^{2}=\vartheta \, q$.
A numerical interpolation starting from the analysis of the first
five peaks for different values of $\vartheta$ has then yielded 
\begin{equation}
\Gamma(\vartheta)=\frac{1}{1+\vartheta}
,
\end{equation}
for which the parametric equation of the lines in Fig.~\ref{cartainst} are given by
\begin{equation}
 \left\{
 \begin{array}{l} 
  q_{n}(\vartheta)=\Gamma(\vartheta)\,n^{2}
  \\
  \\
  \kappa^{2}_{n}(\vartheta)=\vartheta\, \Gamma(\vartheta)\,n^{2}
  .
\end{array}\right.
\end{equation}
On replacing $\vartheta$ from the first into the
second equation, we finally find that the production rate must take its maximum
values at points in the plane $(q,\kappa^2)$ that satisfy the relation
\begin{equation}
\label{k_n_fermi}
 \kappa^{2}_n+q\simeq
 n^{2}
 \ ,
\end{equation}
with $n$ a positive integer.
Note that the above relation represents a very good approximation in the regime
of small oscillations around the static Standard Model Higgs.
\par
The band structure we just described will lead to a preferred production of
non-relativistic particles, that is particles with small momentum compared to their mass.
For example, for fermions with a mass smaller than $M_H/2$, the production is mostly
driven by the first band ($n=1$) and Eq.~\eqref{k_n_fermi} yields a typical momentum
$k^2\sim M_H^2/4-m_f^2$ which is smaller than $m_f^2$ for the particles we consider
in the following.
On the other hand, the production of more massive particles will be caused by higher
order bands (so that $n^2-q>0$) and is normally suppressed.
%
%
%
%
\section{Bosons in a time dependent Higgs \emph{vev}}
\label{analytic-boson}
\setcounter{equation}{0}
The Lagrangian which describes the coupling between the Higgs field and the
vector bosons is the same as in~Eq.\eqref{lagrH}, but with gauge covariant
derivatives and kinetic terms for the gauge fields,
\begin{equation}
\label{lagr-bosoni}
\mathcal{L}_{H-B }=
D^{\mu}H^{\dagger}D_{\mu}H
-\mu^{2}H^{\dagger}H
-\frac{\lambda}{2}\left(H^{\dagger}H\right)^{2}
-\frac{1}{4}F_{\mu\nu}^{i}\,F^{\mu\nu}_{i}
                 -\frac{1}{4}G_{\mu\nu}\,G^{\mu\nu}
,
\end{equation}
where
$D_{\mu}=\partial_{\mu}-ig\frac{\vec{\sigma}}{2}\cdot\vec{A}_{\mu}-i\frac{g'}{2}B_{\mu}$,
$F_{\mu\nu}^{i}=\partial_{\mu}A^{i}_{\nu}-\partial_{\nu}A^{i}_{\mu}
+g\epsilon^{ijk}A^{j}_{\mu}A^{k}_{\nu}$
and
$G_{\mu\nu}=\partial_{\mu}B_{\nu}-\partial_{\nu}B_{\mu}$.
For the electroweak vector bosons we thus have
\begin{eqnarray}
\label{lagrWZ}
 \mathcal{L}_{W^{\pm},Z_{0}}&\!\!=\!\!&
             -\partial_{\nu}W^{+}_{\mu}\partial^{\nu}W^{-\mu}
      -\frac{1}{2}\left(\partial_{\nu}Z_{\mu}\right)^{2} 
      +\frac{g^{2}}{4M^{2}} h^{2}\left(M^{2}W^{+\mu}W^{-}_{\mu}+
            \frac{M_{Z}^{2}}{2}Z^{\mu}Z_{\mu}\right)+
   \nonumber
   \\
  &&+\Phi^{2}\left(M^{2}W^{+\mu}W^{-}_{\mu}
       +\frac{M_{Z}^{2}}{2}Z^{\mu}Z_{\mu}\right)
       +\frac{g}{M}\Phi h \left(M^{2} W^{+\mu}W^{-}_{\mu}
       +\frac{M_{Z}^{2}}{2} Z^{\mu}Z_{\mu}\right).
\end{eqnarray}
Taking the \emph{vev} with the condition $\langle0|h|0\rangle=0$ and neglecting
the \emph{back-reaction} of $h$, the classical equation of
motion for $Z^{\mu}$ is given by
\begin{equation}
 \partial_{\nu}\partial^{\nu}Z^{\mu}+M_{Z}^{2}\Phi^{2}(t)Z^{\mu}=0
.
\end{equation}
The fields $W^{\pm}_{\mu}$ satisfy the same equation with $M_{Z}$ replaced by $M$.
We therefore conclude that any component $\mathscr{Z}$ of the vector field $Z^{\mu}$
and any component $\mathscr{W}$ of $W^{\mp\mu}$ satisfy Klein-Gordon equations
with mass $M_{Z}^{2}(t)\equiv M_{Z}^{2}\Phi^{2}(t)$ and
$M^{2}(t)\equiv M^{2}\Phi^{2}(t)$ which, after performing a spatial Fourier
transform, take the form
\begin{equation} 
\label{eqmodiZ}
 \ddot{\mathscr{Z}}_{k} + \left[k^{2}+M_{Z}^{2}(t)\right]
            \mathscr{Z}_{k}=0\,,
\end{equation}
\begin{equation} \label{eqmodiW}
 \ddot{\mathscr{W}}_{k}+\left[k^{2}+M^{2}(t)\right]
            \mathscr{W}_{k}=0\,,
\end{equation}
where $\mathscr{Z}(t)=(2\pi)^{-3/2}\int d^{3}k\,e^{i\vec{k}\cdot\vec{x}}\mathscr{Z}_{k}(t)$
and $\mathscr{W}(t)=(2\pi)^{-3/2}\int {d^{3}k}\,e^{i\vec{k}\cdot\vec{x}}\mathscr{W}_{k}(t)$.
Analogously, from the Lagrangian~\eqref{lagrH} in the unitary gauge~\eqref{un_g} and keeping
only terms up to second order, one obtains the following equation for the Fourier modes
of the quantum Higgs field
\begin{equation}
\label{h_eq1}
\ddot{{h}}_{k} + \left[k^{2}+M_{h}^{2}(t)\right] {h}_{k}=0
\ ,
\end{equation}
where 
\begin{equation}
M_h^2(t)=M_H^2\left[\frac{3}{2}\,\Phi^2(t)-\frac12\right]
\ .
\end{equation}
\par
We are interested in the case when the Higgs \emph{vev} $\Phi(t)$ oscillates near one
of its two absolute minima (see Fig.~\ref{fig:vacuum_pot}).
We therefore take the limit $m\to +\infty$  ($c \to -1$) in Eq.~\eqref{phi_t} and obtain
\begin{equation}
\Phi(\tau)\simeq 1-\frac{1}{2m}\sin^{2}\tau+O\left(\frac{1}{m^{2}}\right)
\ .
\end{equation}
Moreover, from now on, we shall for simplicity denote with $Y_{\kappa}$ any bosonic mode,
that is $Y_{\kappa}$ can be either $\mathscr{Z}_{\kappa}$, $\mathscr{W}_{\kappa}$ or
$h_\kappa$.
Eqs.~\eqref{eqmodiZ}, \eqref{eqmodiW} and \eqref{h_eq1} then become
\begin{eqnarray}
\label{eq_modi_bos}
Y''_{\kappa}(\tau) +\varpi_\kappa^2(\tau)Y_{\kappa}(\tau)
=0
\ ,
\end{eqnarray}
with
\begin{eqnarray}
\label{om_b}
\varpi_\kappa^2(\tau)=\kappa^{2}+q\left(1-\frac{\sigma_Y}{m}\sin^{2}\tau\right)
\ ,
\end{eqnarray}
where we have neglected terms beyond the first order in ${1}/{m}$ and introduced
the dimensionless constants
\begin{equation}
\label{boso_par}
\kappa\equiv\frac{2k}{M_{H}}
\quad
{\rm and}
\quad
q\equiv 4\frac{M_{Y}^{2}}{M_{H}^{2}}
\ ,
\end{equation}
with $M_Y=M_Z$, $M$ or $M_H$ and $\sigma_Y=1$ for the gauge fields and $3/2$
for the Higgs.
Note that Eq.~\eqref{eq_modi_adim} for the fermion modes takes the same form as
Eq.~\eqref{eq_modi_bos} when the imaginary part of the fermion frequency can be
neglected and that the period of $\varpi_\kappa(\tau)$ is $\pi$.
\par
If we define
\begin{equation}
\label{a-epsilon}
  a\equiv \frac{1}{4}\left[\kappa^{2}
           +q\left(1-\frac{\sigma_Y}{2m}\right)\right]
           \ ,
   \qquad
     \epsilon \equiv \frac{q\,\sigma_Y}{16\,m}
  \ ,
\end{equation}
and change the time to $\eta \equiv 2\tau=M_{H}t$, we obtain
\begin{equation}
\label{mathieu}
 Y''_{\kappa}(\eta)
  +\left(a+2\epsilon\cos\eta\right) Y_{\kappa}(\eta)=0
  \ .
\end{equation}
When the total vacuum energy $c$ is near the minimum value $c=-1$
(for which $\Phi$ is constant), $W^{\pm\mu}$, $Z^{\mu}$ and $h$ therefore satisfy a
\emph{Mathieu equation}.
\par
\begin{figure}[t]
   \centering
   {\includegraphics[scale=0.80]{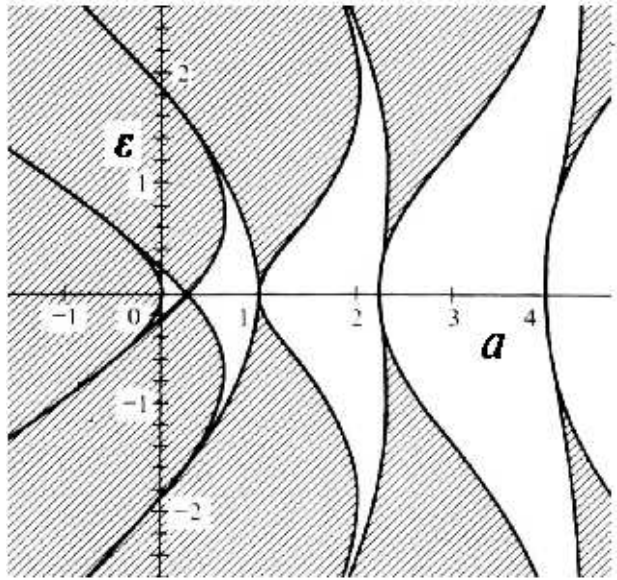}}
 \caption{Stability regions for the Mathieu equation.
Stable solutions have $a$ and $\epsilon$ in the unshaded regions.
When $\epsilon=0$, instability bands cross the $a$-axis at $a_{n}=n^{2}/4$,
with $n=0,1,2,\ldots$.}
\label{Mcarta-inst}
\end{figure}

The Mathieu equation is a linear differential equation with time periodic
coefficients and is described in the general framework of Floquet theory
\cite{bender_orszag}.
An important result from this theory is the existence of stable solutions only
for particular values of $a$ and $\epsilon$  (see Fig.~\ref{Mcarta-inst}).
For our purposes, the relevant solutions which lead to an efficient particle
production in the quantum theory are however those which show an exponential
instability of the form ($\mu_{\kappa}^{(n)}$ is known as Floquet index or
growth factor)
\begin{equation}
\label{floquet_ind}
Y_{\kappa} \sim \exp\left(\mu_{\kappa}^{(n)}\eta/2\right)
\end{equation}
and appear within the set of resonance bands of width $\Delta \kappa^{(n)}$ labelled
by the integer index $n$.
Using Eq.~\eqref{a-epsilon} one can then map the stability chart of
Fig.~\ref{Mcarta-inst} into the plane $(q,\kappa^2)$ and finds that the instabilities
also correspond to an exponential growth of the occupation number of quantum
fluctuations, $n_{\kappa}\propto\exp(\mu_{\kappa}^{(n)}\eta)$ (see
Eq.~\eqref{inviluppo_n(t)_bos} below), which can be interpreted as strong
particle production.
Stable solutions also lead to particle production, although with no exponential
growth in time, and thus resemble the fermion case \cite{mostepanenko_frolov}.
%
\par
If $c\approx-1$ then $m\gg1$ and $\epsilon\approx0$ in Fig~\ref{Mcarta-inst}.
One then finds $4\,a\approx n^{2}$, or
\begin{equation}
\label{k+q=n}
  \kappa^{2}_n+q\left(1-\frac{\sigma_Y}{2m}\right)=n^{2}\ , \qquad n=1,2,3,\ldots
  \ ,
\end{equation}     
which, for a given $q$, gives the value of $\kappa^{2}$ around which the $n$-th
resonance peack is centred.
Further, the physical constraint $\kappa^{2} \geq 0$ implies $q \leq 2\,n^{2}\,m/(2m-1)$,
so if $q \lesssim 1$ ($M_{H} \gtrsim 2M_{Y}$) all the resonance bands contribute to
production, otherwise at least the first band ($n=1$) is not available.
In particular, the first band is never available for the Higgs particle and its production can
therefore be negligible with respect to that of gauge bosons if the Higgs mass is
sufficiently large. 
%
%
%
\subsection{Boson occupation number and energy density}
%
The same analytical steps of Section~\ref{sec-fermion_theory}
allows one to relate relevant physical quantities to the solutions of the
equation of motion also in the case of bosons.
We again take initial conditions corresponding to positive frequency
plane waves for $\tau\leq 0$,
\begin{equation}
\label{cond_iniz_bos}
\left\{
\begin{array}{l}
 Y_{\kappa}(0)=\left[{2\varpi_{\kappa}(0)}\right]^{-1/2}
  \\
  \\ 
 Y_{\kappa}'(0)=-i\varpi_{\kappa}(0)Y_{\kappa}(0)
 .
\end{array}
\right.
\end{equation}
The occupation numbers for bosons will then be given by
\cite{Mostepanenko_book}
\begin{equation}
\label{no_bos_esatto}
 n_{\kappa}(\tau)=\frac{1}{2\varpi_{\kappa}}|Y_{\kappa}'|^{2}
                  +\frac{\varpi_{\kappa}}{2}|Y_{\kappa}|^{2}-\frac{1}{2}
                  \ ,
\end{equation}
where $\varpi_{\kappa}=\varpi_{\kappa}(0)$ and note that $n_{\kappa}(0)=0$
thanks to eqs.~\eqref{cond_iniz_bos}. 
\par
Like for fermions, it is possible to find an analytic approximation for the boson occupation
number  \cite{Mostepanenko_book},
\begin{equation}
\label{inviluppo_n(t)_bos}
 \hat{n}_{\kappa}=\sinh^{2}(\mu_{\kappa}\tau)\,,
\end{equation}
with the \emph{Floquet index}, or growth factor, $\mu_{\kappa}$ given by
\begin{equation}
\label{floquetindex}
 \cosh (\mu_{\kappa} T)=\re \left[Y_{\kappa}^{(1)}(T)\right]
 \ ,
\end{equation}
in which $Y_{\kappa}^{(1)}$ is the solution of Eq.~\eqref{eq_modi_bos}
with initial conditions $Y_{\kappa}^{(1)}(0)=1$,
$Y_{\kappa}^{(1)'}(0)=0$ and $T=\pi$ is the period of
$\varpi_{\kappa}(\tau)$ (the period of the Higgs vacuum). 
Using this approximate expression, the boson energy density can be written as
\begin{equation}
\label{densita_en_bos}
 \tilde{\rho}_{B}(\tau)=\frac{1}{2\pi^{2}}\int d\kappa\, \kappa^{2} 
     \varpi_{\kappa}(\tau)\sinh^{2}(\mu_{\kappa}\tau)
     \ ,
\end{equation}  
where the tilde on $\rho$ is to remind that this is a
dimensionless quantity (like the fermion analogue in Eq.~\eqref{densita_energia_adim}).
%
%
%
%
\section{Fermion production}
\label{fermi_pro}
\setcounter{equation}{0}
We shall now use the numerical solutions to Eq.~\eqref{eq_modi_adim} to evaluate
the fermion occupation number $n_{\kappa}$ in Eq.~\eqref{numero_occupazione_adim} and
compare it with its analytical approximation $\hat n_{\kappa}$ in Eq.~\eqref{inviluppo_n(t)_adim}.
In order to illustrate the magnitude of the effect, we shall consider three
possible values for the Higgs boson mass which result from different experimental
lower or upper bounds, namely $M_H=115\,$GeV, $200\,$GeV and
$500\,$GeV~\cite{higgs-mass}.
In this Section, we shall not take into account
the back-reaction of the produced fermions which is treated later.
\begin{figure}[t] 
   \subfigure
   {\includegraphics[scale=0.83]{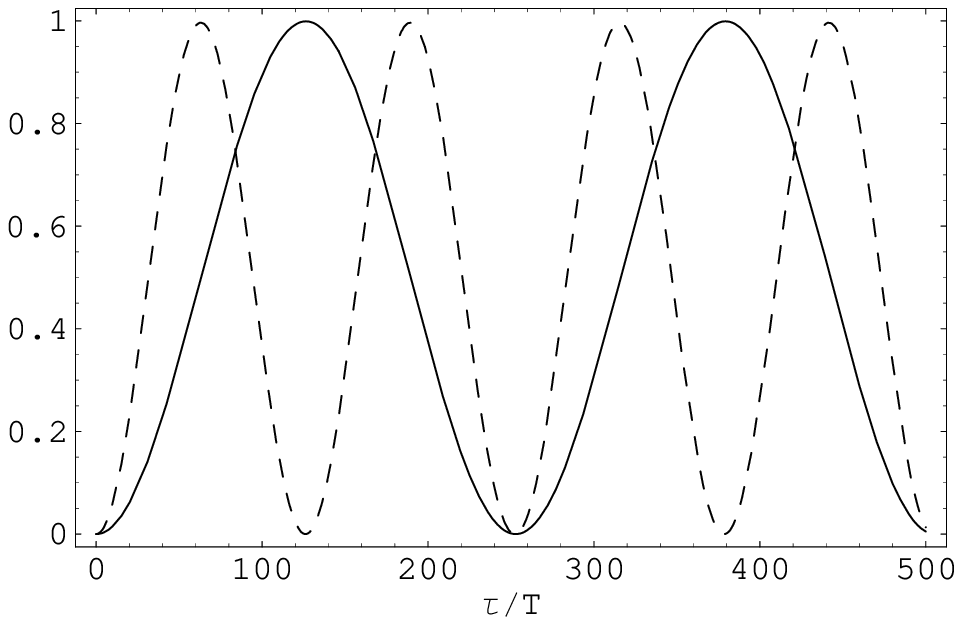}}
   \subfigure 
   {\includegraphics[scale=0.78]{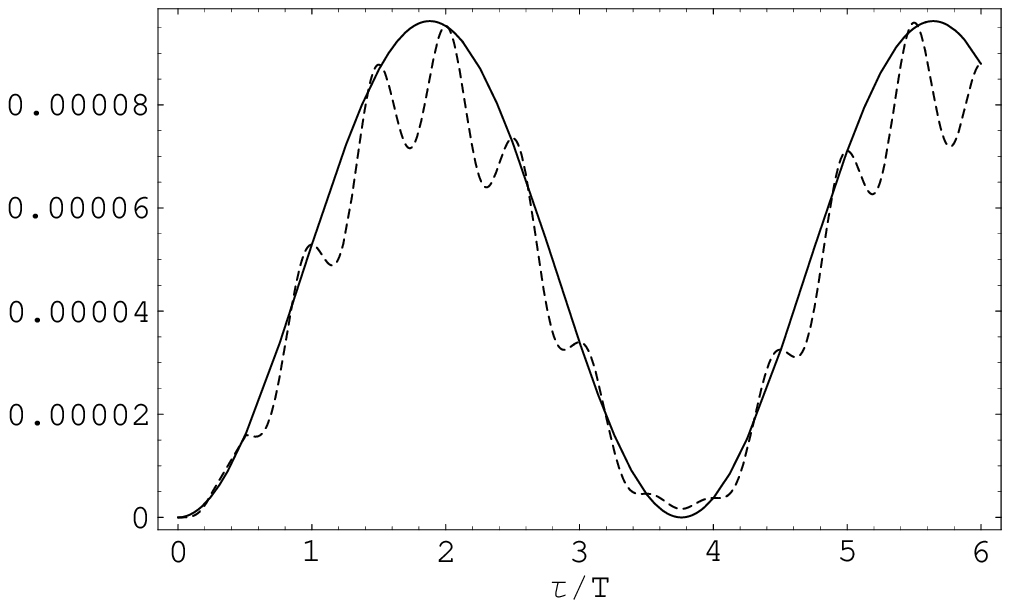}}    
 \caption{Time evolution of the occupation number $n_{\kappa}$ and its
   envelope $\hat{n}_{\kappa}$ for the \emph{top} quark and $M_H=500\,$GeV.
   Left graph: $\kappa^2=1-q\simeq 0.51$ on the first resonance band,
   $c_0=-1+10^{-3}$ (solid line) and
   $c=-1+4\cdot 10^{-3}$ (dashed line).
   Note that $n_{\kappa}$ and $\hat{n}_{\kappa}$ exactly coincide and their
   period scales as $(c+1)^{1/2}$. 
   Right graph: $\kappa^{2}=q/10\simeq 4.8\cdot 10^{-2}$ is outside resonance bands and
   $\hat{n}_{\kappa}$ (solid line) differs from $n_{\kappa}$ (dashed line).}
\label{fig:NO&detNO}
\end{figure}
%
\begin{figure}[t] 
   \subfigure
   {\includegraphics[scale=0.83]{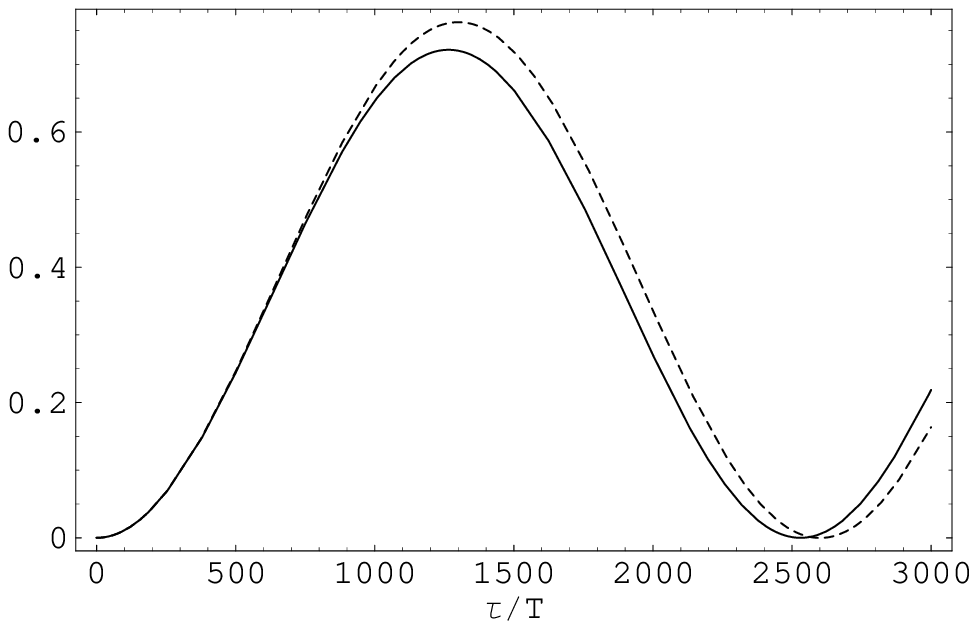}}
   \subfigure 
   {\includegraphics[scale=0.78]{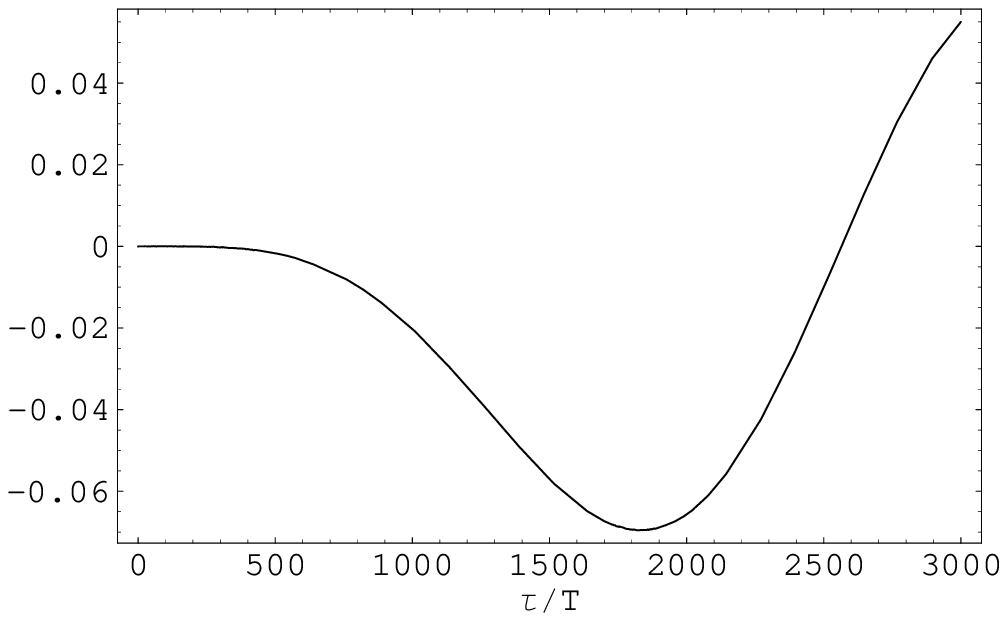}}    
 \caption{Time evolution of the occupation number for the \emph{bottom} quark, $M_H=200\,$GeV,
  $\kappa^2=1-q\simeq 0.9982$ on the first resonance band and $c_0=-1+10^{-3}$.
  Left graph: $n_{\kappa}$  (dashed line) and its
   envelope $\hat{n}_{\kappa}$  (solid line).
   Right graph: difference between $n_{\kappa}$ and $\hat{n}_{\kappa}$.}
\label{fig:NObot}
\end{figure}
%
\par
Let us begin with Fig.~\ref{fig:NO&detNO}, which shows the time evolution of the occupation
number $n_{\kappa}$ and its enveloping $\hat{n}_{\kappa}$ for the \emph{top}
quark in the case $M_{H}=500\,$GeV and two different values of the momentum $\kappa$,
and Fig.~\ref{fig:NObot}, which shows the same quantities for the \emph{bottom}
quark in the case $M_{H}=200\,$GeV:
\begin{description}
\item[a)]
For $\kappa$ on the (first) resonance band both functions $n_{\kappa}$ and $\hat{n}_{\kappa}$
for the \emph{top} reach the maximum allowed by the Pauli blocking and are remarkably indistinguishable (left graph in Fig.~\ref{fig:NO&detNO}).
In fact, their difference remains of the order of $10^{-8}$ for both values of the
initial background energy $c=c(\tau=0)$ and we do not show it.
For the \emph{bottom} (\emph{i.e.}, a smaller value of $q$ at the resonance with respect to
the \emph{top}'s) instead, the two
functions differ slightly and always remain smaller than one (see Fig.~\ref{fig:NObot}).
In both cases however, from the numerical simulations one can infer a \emph{scaling law}
for the period $T_\kappa$ of the occupation number with respect to the initial vacuum energy
$c$.
If $c_0$ is a reference energy, one has
\begin{equation}
T_\kappa(c)=T_\kappa(c_0)\,\left(\frac{c_0+1}{c+1}\right)^{n/2}
\ ,
\end{equation}
which holds for all values of $\kappa$ on the $n$-th resonance band
provided $c$ is small enough that only small oscillation of the background $\Phi$
are relevant (see case~\ref{-1c0} in Section~\ref{higgs_t}).
Note also that $T_\kappa\gg T$ for the \emph{top} (we recall that the Higgs \emph{vev}
period is $2T/M_H \sim 10^{-26}\,$s).
\item[b)]
For values of $\kappa$ not on a resonance band (right graph in Fig.~\ref{fig:NO&detNO}),
the production
is of course much damped and the exact occupation number shows relatively
high frequency oscillations, with
period comparable to the Higgs \emph{vev}'s,
which are modulated by the function $\hat n_\kappa$ with a period usually larger
than $T$.
It is therefore clear that the time average over these relatively high frequency
oscillations can still be related to the enveloping $\hat n_\kappa$ as
\begin{equation}
\label{mean_n_k}
\bar n_\kappa(\tau)=
\langle n_{\kappa}(\tau)\rangle_{\tau}\equiv
\frac{1}{T}\int_{\tau}^{\tau+T}\!d\zeta\,n_{\kappa}(\zeta)
\simeq
\hat{n}_{\kappa}(\tau)
\ ,
\end{equation}
where $\hat{n}_{\kappa}$ is given in Eq.~\eqref{inviluppo_n(t)_adim}.
\end{description}
We emphasize that the fermion occupation number is always periodic in time
and the function in Eq.~\eqref{inviluppo_n(t)_adim} \emph{exactly} equals
$n_{\kappa}(pT)$, with $p$ a positive integer, regardless of the fact that
$\kappa$ and $q$ are on a resonance band or not (see Fig.~\ref{fig:NO&detNO}).
This is a general behaviour which still holds for very small values of $q$,
for example $q\sim 10^{-10}$ or $q\sim 10^{-22}$, corresponding
to the scales of the electron and neutrino mass ($\lesssim 1\,$eV) respectively.
These cases are however very difficult to treat numerically because one should
integrate for very long times (the period of the function in Eq.~\eqref{inviluppo_n(t)_adim}
grows with the inverse of the fermion mass).
For this reason we have chosen to show the occupation number only for
quarks like the \emph{top} and \emph{bottom}.
\par
We are now interested in the production probability for a given fermion, {\em i.e.}, in the
occupation number $n_\kappa$ summed over all $\kappa$'s for
a given value of $q$ (see Eq.~\eqref{parametri_adim}).
The generally oscillating behaviour of $n_\kappa$ leads us to estimate this probability
in time as the mean number \eqref{mean_n_k} summed over all the momenta $\kappa$,
\begin{equation}
\bar{n}_{\psi}
\equiv \langle n_{\psi}(\tau)\rangle_{\tau}
= \frac{1}{4\,\pi^{3}}\int \!d^{3}\kappa\,F_{\kappa}
 \langle\sin^{2}(\nu_{\kappa}\tau)\rangle_{\tau}
=\frac{1}{4\,\pi^{3}}\int\! d^{3}\kappa\,\frac{F_{\kappa}}{2}
\ .
\end{equation}
As we see from Fig.~\ref{picchi(M_H)}, the mean occupation number $\bar n_\kappa$ as a
function of $\kappa$ at fixed $q$ is significantly different from zero only near the values
$\kappa^{2}=\kappa_n^2$ around which the production peaks are centred.
We can therefore estimate $\bar n_\psi$ as
\begin{equation}
\label{prob_di_prod_media}
\bar n_\psi \simeq \frac{1}{4\,\pi^{3}} \sum_{n=1}^{n_{p}}
         \int\displaylimits_{Supp(P_{n})}\!\!\!\! d^{3}\kappa\,\bar{n}_{\kappa}
         \simeq
         \frac{1}{2\,\pi^{2}} \sum_{n=1}^{n_{p}} 
         \sqrt{\kappa_n^2}\,\Delta\kappa^2_n\,\bar{n}_{\kappa_n}
\end{equation}
where $Supp(P_{n})$ is the interval around $\kappa^{2}_n$ in which $\bar{n}_{\kappa}$
is significantly large.
The effective width of the peak $\Delta\kappa^2_n$ on the $n$-th band
is determined by estimating each integral in the above expression numerically.
Further, we found that
\begin{equation}
\label{DKscale}
\Delta\kappa^2_n(c)\simeq
\Delta\kappa^2_n(c_0)\,\left(\frac{c+1}{c_0+1}\right)^{\frac{n}{2}}
\ ,
\end{equation}
a scaling behaviour also shared by the bosons (see Section~\ref{boso_pro}).
\par
%
\begin{table}[t]
\centering
 \begin{tabular}{|c||c||c|}
 \hline 
 \multicolumn{3}{|c|}
 {$\mathbf{M_{H}=115\,GeV}$}\\
 \hline \hline
  $\psi$ & $\mathbf{\bar n_{\psi}}$ & $q$ \\
 \hline \hline
  $t$    & $\ll 10^{-10}$      & $9.26$ \\ 
 \hline
  $b$    & $8.7 \cdot 10^{-5}$ & $5.46 \cdot 10^{-3}$ \\
 \hline
  $d$    & $1.6 \cdot 10^{-7}$ & $1.38 \cdot 10^{-8}$ \\
 \hline
  $\tau$ & $3.9 \cdot 10^{-5}$ & $9.54 \cdot 10^{-4}$ \\
 \hline
 \end{tabular}
 \begin{tabular}{|c||c||c|}
 \hline 
 \multicolumn{3}{|c|}
 {$\mathbf{M_{H}=200\,GeV}$}\\
 \hline \hline
  $\psi$ & $\mathbf{\bar n_{\psi}}$ & $q$ \\
 \hline \hline
  $t$    & $1.6 \cdot 10^{-5}$ & $3.06$  \\ 
 \hline
  $b$    & $5.1 \cdot 10^{-5}$ & $1.80 \cdot 10^{-3}$ \\
 \hline
  $d$    & $8.2 \cdot 10^{-8}$ & $4.54 \cdot 10^{-9}$ \\
 \hline
  $\tau$ & $2.4 \cdot 10^{-5}$ & $3.16 \cdot 10^{-4}$ \\
 \hline
 \end{tabular}
\centering 
 \begin{tabular}{|c||c||c|}
 \hline 
 \multicolumn{3}{|c|}
 {$\mathbf{M_{H}=500\,GeV}$}
 \\
 \hline \hline
  $\psi$ & $\mathbf{\bar n_{\psi}}$  & $q$\\
 \hline \hline
  $t$    & $4.3 \cdot 10^{-4}$  & $0.48$\\ 
 \hline
  $b$    & $2.4 \cdot 10^{-5}$  & $2.90 \cdot 10^{-4}$\\
 \hline
  $d$    & $3.3 \cdot 10^{-8}$  & $7.28 \cdot 10^{-10}$ \\
 \hline
  $\tau$ & $8.8 \cdot 10^{-6}$  & $5.06 \cdot 10^{-5}$\\
 \hline
 \end{tabular}
 \caption{Production probability of various quarks and of the lepton
 \emph{tau} for three different values of the Higgs mass and
Higgs total vacuum energy $c_0=-1+10^{-3}$.}  
 \label{tabelle_prob_di_prod}
\end{table}
%
%
%
The occupation number in Eq.~\eqref{prob_di_prod_media} for the \emph{top}, \emph{bottom},
\emph{down} and for the \emph{tau} are shown in Table~\ref{tabelle_prob_di_prod} for three
different values of the Higgs mass and Higgs total vacuum energy $c_0=-1+10^{-3}$.
Note that the production is mostly generated from the first band and,
if $q<1$, the production probability is larger than for
$q>1$ since for the latter case only the bands of order greater than one can
contribute.
So, recalling that
$q=4\,{m_{f}^{2}}/{M_{H}^{2}}$, we conclude that in the Standard
Model all the fermions except the \emph{top} have a greater production
probability for a ``light'' Higgs.
Moreover, within a given band, the production is still strongly affected by
the resonance parameter $q$.
On varying $q$ along a given band one observes a
change in the value of the momentum $\kappa$ of the produced particles
(see Eq.~\eqref{k_n_fermi}), in the width of the band and, consequently, in
the number density of produced particles (see Figs.~\ref{cartainst} and
\ref{picchi(M_H)}).
These facts lead to the values of $\bar n_\psi$ (as defined in
Eq.~\eqref{prob_di_prod_media} above) presented in Table~\ref{tabelle_prob_di_prod}.   
Note finally that the \emph{bottom} is the
``dominant'' (with the highest production probability) fermion
for $M_{H}=115\,$GeV and $M_{H}=200\,$GeV, while for
$M_{H}=500\,$GeV the \emph{top} production is more probable.
\par
We remark that the average occupation number $\bar n_{\psi}$
also scales with the energy according to the relation
\begin{equation}
\label{n(energy)}
\bar n_{\psi}(c)\simeq
n_{\psi}(c_0)\left(\frac{1+c}{1+c_0}\right)^{\frac{n}{2}}
\sim(1+c)^{\frac{n}{2}}\,, \qquad |c+1| \lesssim 10^{-3}
 \ ,
\end{equation}
where $n$ is the order of the ``dominant'' peak (the one which contributes most to the production).
So if one is interested in the production probabilities at an arbitrary but
always very small energy, the values in Table~\ref{tabelle_prob_di_prod},
calculated for $c_0=-1+10^{-3}$,
must be multiplied by the factor $[(1+c)/(1+c_0)]^{\frac{1}{2}}$ if the
production is associated with the first band.
For example, this is true for the {\it top} if $M_H> 2\,m_{\rm top}$ and all other fermions
for any Higgs mass.
Instead, in the region $M_H< 2\,m_{\rm top}$ (of phenomenological interest),
the {\it top} may be produced only starting from the second band and the relevant scale
factor is $(1+c)/(1+c_0)$.
\par
Finally, we evaluate the fermion energy density using Eq.~\eqref{densita_energia_adim}.
The integral over $\kappa$ can be calculated in the same way as in
Eq.~\eqref{prob_di_prod_media} and including just the dominant peak for the production.
This approximation is better the lower the Higgs vacuum energy,
since the peak amplitude decreases proportionally to this energy.
we thus find
\begin{equation}
\label{dens-en-connu}
 \tilde{\rho}_{\psi}(\tau)=\frac{1}{2\pi^{3}}\!\!
                        \int \! d^{3}\kappa\, \omega_{\kappa}(\tau)
                            \frac{F_{\kappa}}{2} \sin^{2}(\nu_{\kappa}\tau)
                        \simeq \frac{1}{2\pi^{3}}\!\!\!\!\!\!\!\!\!\!
                               \int\displaylimits_{Supp(P_{dom})}
                               \!\!\!\!\!\!\!\!\!
                                  d^{3}\kappa \, 
                                \omega_{\kappa}(\tau)\,
                                \bar{n}_{\kappa} \sin^{2}(\nu_{\kappa}\tau)
                                \ . 
\end{equation}
Around the dominant peaks $\bar{\kappa}_n^{2}$ given by Eq.~\eqref{k_n_fermi}
we have
\begin{equation}
 \nu_{\bar{\kappa}_{n}^{2}}\approx 1
\ ,
\end{equation}
and we can therefore write the energy density \eqref{dens-en-connu} as
\begin{equation}
\label{dens-en_appross}
 \tilde{\rho}_{\psi}(\tau)\simeq \frac{\sin^{2}(\tau)}{2\,\pi^{3}}\,
                                     \!\!\!\!\!\!\!\!\!    
                               \int\displaylimits_{Supp(P_{dom})}
                                      \!\!\!\!\!\!\!\!\!
                                d^{3}\kappa \; 
                                \omega_{\kappa}(\tau)\,
                                \bar{n}_{\kappa}
                                \simeq
                                \frac{\sin^2(\tau)}{\pi^2}\sum_{n=1}^{n_p}
                                \sqrt{\kappa_n^2}\,\Delta \kappa^2_n\, 
                                \omega_{\kappa_n}(\tau)\,{\bar n_{\kappa_n}}                            
                                \ ,
\end{equation}
which significantly simplifies the evaluation of the integral over $\kappa$. 
One can also find a scaling law for the energy density so obtained, namely
\begin{equation}
\tilde{\rho}_{\psi} \sim (1+c)^{\frac{n}{2}}
\ ,
\end{equation}
with the same prescriptions as for $n_{\psi}$.
\par
\begin{figure}[t]
   \centering
   {\includegraphics[scale=0.8]{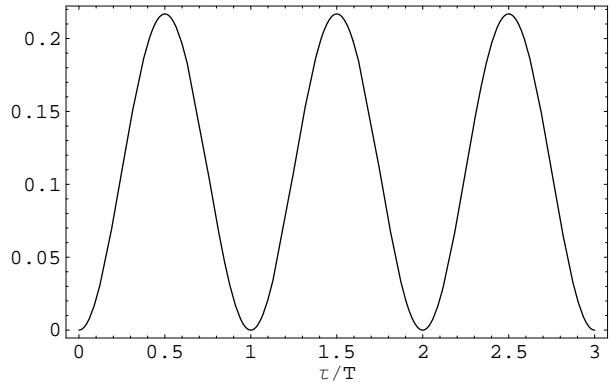}}
 \caption{Fraction of Higgs vacuum energy $R$ absorbed by the produced \emph{top}
 for $M_H=500\,$GeV and $c_0=-1+10^{-3}$.}
\label{plot_densEnFer}
\end{figure}

From Eq~\eqref{lagr_phi,h}, the vacuum energy in a volume
${8}/{M_{H}^{3}}$ can be calculated and put in the usual dimensionless form
as
\begin{equation} 
 \widetilde{\mathcal{E}}_{\Phi}\equiv \frac{16}{M_{H}^{4}}\mathcal{E}_{\Phi}
                    =\frac{2}{\lambda}\,c
                    \ .
\end{equation}
The Higgs vacuum energy density available for the production is thus given by
\begin{equation}
\label{deltaEHiggs}
  \Delta \widetilde{\mathcal{E}}_{\Phi}
  \equiv \widetilde{\mathcal{E}}_{\Phi}-\widetilde{\mathcal{E}}_{\Phi=1}
     =\frac{2}{\lambda}\,(c+1)
     \ .
\end{equation}
In the Standard Model $\lambda$ appears as an arbitrary parameter and
there are presently no experimental constraints on it.
There are however good reasons to believe that $\lambda < 1$ but we shall
consider in all our calculations simply the value $\lambda = 1$.
Taking into account the dependence of
$\langle \tilde{\rho}_{\psi}\rangle$ and
$\Delta \widetilde{\mathcal{E}}_{\Phi}$ on $c$,
the fraction of total energy absorbed by a given fermion scales as
\begin{equation}
\label{R}
R(c)=\frac{\langle \tilde{\rho}_{\psi}\rangle}{\Delta \widetilde{\mathcal{E}}_{\Phi}}
=R(c_0,\lambda)\,\left(\frac{1+c_{0}}{1+c}\right)^{n/2} \,, \quad
R(c_0,\lambda)\equiv \frac{\lambda}{2}(1+c_0)^{-n/2} \langle \tilde{\rho}_{\psi}(c_0)\rangle
\end{equation}
An example of this quantity is plotted in Fig.~\ref{plot_densEnFer}.
Notably, the maximum value is reached before the end of the first vacuum oscillation.
%
%
\begin{table}[t]
\centering
 \begin{tabular}{|c||c||c|c|}
 \hline
  $\psi$ & $R$  & $q$ & $M_H$\\
 \hline \hline
  $t$    & $1.1 \cdot 10^{-1}$ & $0.48$ & $500\,$GeV\\ 
 \hline
  $b$    & $2.2 \cdot 10^{-2}$ & $5.46 \cdot 10^{-3}$ & $115\,$GeV\\
 \hline
  $d$    & $4.0 \cdot 10^{-5}$ & $1.38 \cdot 10^{-8}$ & $115\,$GeV\\
 \hline
  $\tau$ & $1.0 \cdot 10^{-2}$ & $9.54 \cdot 10^{-4}$ & $115\,$GeV\\
 \hline
 \end{tabular}
 \caption{Fraction of Higgs vacuum energy absorbed by various quarks and the \emph{tau} lepton
 for the particular Higgs mass (out of the three considered in Table~\ref{tabelle_prob_di_prod}
 with $c_0=-1+10^{-3}$)
 which gives the highest production rate.}  
 \label{tabella_dens_en}
\end{table}
\par
Using the results shown in the Table~\ref{tabelle_prob_di_prod}, for each
included fermion we have selected the values of the Higgs mass which give the greatest
production probability and the corresponding energy densities are shown in
Table~\ref{tabella_dens_en} in units of the Higgs vacuum energy density avaible
for the production $\Delta\widetilde{\mathcal{E}}_{\Phi}$.
Of course, the physical condition $R<1$ limits the range of
validity of this first approximation where production is not taken to affect the
{\it vev} oscillation (no back-reaction), breaking energy conservation.
For example, the value of $R$ given in Table~\ref{tabella_dens_en} 
for the \emph{top} for $c_0=-1+10^{-3}$ can be rescaled down to a minimum energy
$c\simeq -1+4\cdot 10^{-5}$, for our choice of $\lambda$.
\par
When their energy density becomes comparable with
the background energy, the produced fermions are expected to back-react on the
Higgs \emph{vev}, thus affecting its evolution and eventually suppressing
the parametric production of particles.
We shall see this in detail in Section~\ref{back}.
For now we just note that the different scaling law for the \emph{top}
quark energy density as a function of $c$
(when the Higgs mass is large) implies that back-reaction
effects will become important sooner and
one should therefore be aware that in this case, for very small Higgs energy
oscillations, the \emph{top} energy density produced will be strongly affected
by these effects.
\par
%
%
%
%
\section{Boson production}
\label{boso_pro}
\begin{figure}[t]
   \centering
  {\includegraphics[scale=1.1]{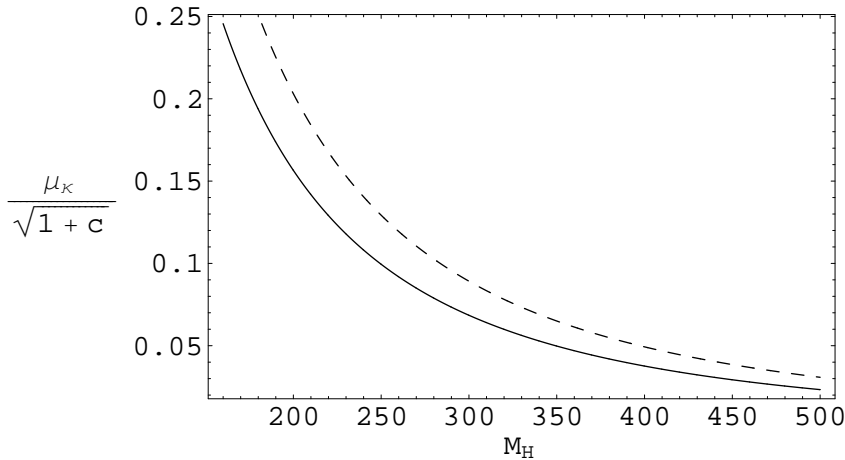}}
 \caption{Floquet index for the first resonance band $\mu_{\kappa}^{(1)}$
 as a function of the Higgs mass for $Z^0$ (dashed line) and $W^\pm$
(solid line).
 In both cases $\mu_{\kappa}^{(1)}$ is maximum if $M_{H}\approx 2\,M_{Y}$
 and decreases for larger values.
 A factor $\sqrt{1+c}$ on the vertical axis is used in consideration of
Eq.~\eqref{Flo_num} with $c_{0}=-1+10^{-3}$.}
\label{plot_floquetindex}
\end{figure}

\setcounter{equation}{0}
In Section~\ref{analytic-boson} we reviewed the fact that the equation of
motion for a bosonic field admits unstable solutions only for some values of
the parameters $q$ and $\kappa$.
The results of the numeric integration of Eq.~\eqref{eq_modi_bos} will be used to
evaluate the occupation number for every mode $\kappa_{n}$ given in
Eq.~\eqref{k+q=n}.
For both $Z_{0}$ and $W^{\pm}$ we have seen that, for a given
value of $M_{H}$, the first band ($n=1$) contributes to the production in an
extremely dominant way with respect to the others
since the Floquet
numbers scale with the Higgs vacuum energy according to
\begin{equation}
\mu_{\kappa}^{(n)}(c)\simeq
\left(\frac{1+c}{1+c_0}\right)^{\frac{n}{2}}\mu_{\kappa}^{(n)}(c_{0})
\ ,
\label{Flo_num}
\end{equation}
where we usually take $c_{0}=-1+10^{-3}$.
Therefore, in the following we shall just refer to the first band and then
do not describe the Higgs production (see Section~\ref{analytic-boson}),
which could be trivially included.
Moreover, motivated by the discussion at the end of Section~\ref{analytic-boson},
we shall first study the case $M_{H}\simeq2\,M_{Z^{0}}=182\,$GeV as the plot in
Fig.~\ref{plot_floquetindex} shows that this yields a more efficient production.
\par
Let us note that for the
case of production dominated by the second band (\emph{i.e.}, for bosons with mass less
than twice the Higgs mass, including the Higgs itself), the relevant
Floquet index is much smaller.
In particular, a plot similar to that in Fig.~\ref{plot_floquetindex}
would display a curve for $\mu_{\kappa}^{(2)}$, which scales as $(1+c)$ instead
of $(1+c)^{1/2}$, with maximum value on the vertical axis of the order of $10^{-2}$.
%
%
\begin{figure}[t]
   \centering
   \subfigure
   {\includegraphics[scale=0.8]{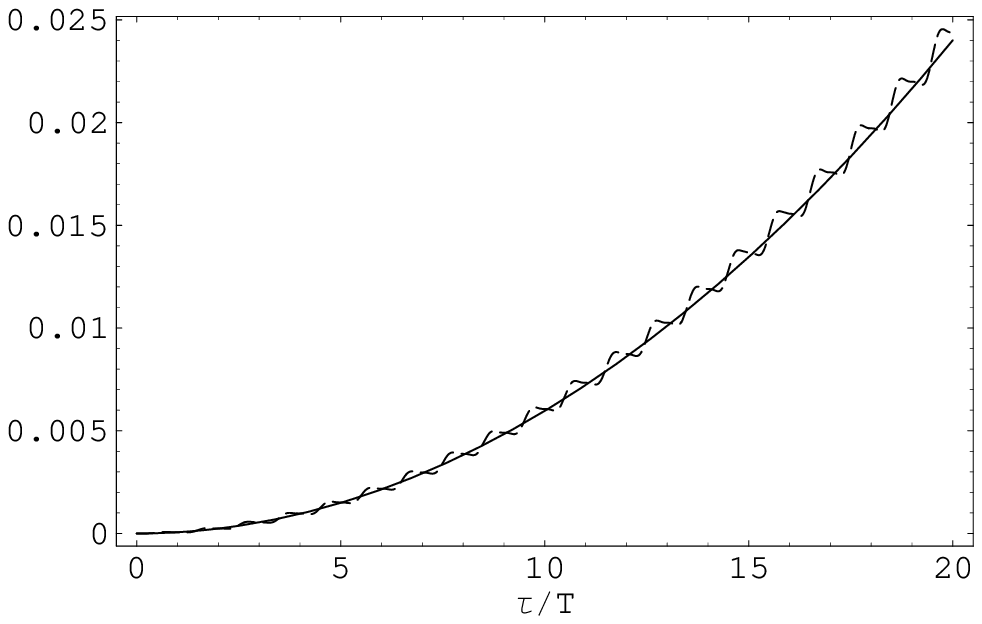}}
    \subfigure
   {\includegraphics[scale=0.8]{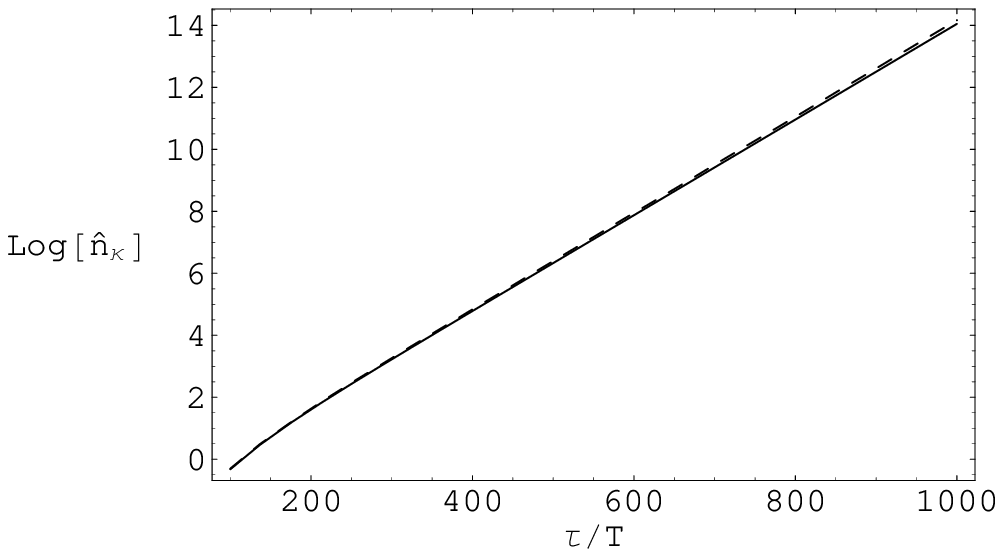}}
 \caption{Time evolution of the number of produced $Z^0$ with
 $\kappa^2=\kappa_1^2\simeq 0.17$ for two time ranges,
 $M_H=200\,$GeV and $c_0=-1+10^{-3}$.
 Note that $\hat n_\kappa$ (solid line) matches $n_\kappa$ (dashed line)
 at the end of each background oscillation.}
\label{plot_nk_Bos}
\end{figure}
\par
Fig.~\ref{plot_nk_Bos} shows the time dependence of the occupation number 
\eqref{no_bos_esatto} and its enveloping approximation \eqref{inviluppo_n(t)_bos}
for the boson $Z^0$ with a momentum on the first resonance band,
$\kappa^2=\kappa_1^2\simeq 0.17$, $M_H=200\,$GeV and $c_0=-1+10^{-3}$.
The exact $n_\kappa$, whose mean value grows exponentially with time, oscillates and coincides
with $\hat n_\kappa$ at the end of each background oscillation.
\par
To evaluate the production probability (\emph{i.e.}, the total occupation number) for a given
boson we must integrate the mode occupation number $n_\kappa$
over a finite volume in momentum space (at fixed $q$).
This very difficult integration is greatly simplified by the fact that the production
only arises around the peaks in the $(q,\kappa^2)$ plane, that is for
$\kappa^2=\kappa^2_n$ given in Eq.~\eqref{k+q=n}.
Moreover, the production from the first instability band is the most relevant,
so that the integration can be consistently restricted to an interval
around $\kappa_{1}$ (denoted as $Supp({P_{1}})$, like for fermions),
\begin{equation}
\label{BON_approx}
n_{B}(\tau)=\frac{1}{(2\,\pi)^{3}}\int d^{3}\kappa\,\hat{n}_{\kappa}(\tau)
      \approx \frac{1}{4\,\pi^{2}}
\!\!\!\!\!\!\!\int\displaylimits_{Supp(P_{1})}\!\!\!\!\!\!\!\! d\kappa^{2}\,
\sqrt{\kappa^2}\sinh^2\left({\mu_{\kappa}\tau}\right)
\ .
\end{equation}
We shall then consider two different time scales.
For short times ($\mu_{\kappa}\tau\ll 1$),
the above expression can be approximated as
\begin{equation}
\label{BON_approx_s}
n_{B}(\tau)\approx \frac{1}{2\,\pi^{2}}
\!\!\!\!\!\!\!\int\displaylimits_{Supp(P_{1})}\!\!\!\!\!\!\!\! d\kappa^{2}\,
\sqrt{\kappa^2}\left({\mu_{\kappa}\tau}\right)^2
\ ,
\end{equation}
whereas for long times ($\mu_{\kappa}\tau\gg 1$) we shall use
\begin{equation}
\label{BON_approx_l}
n_{B}(\tau)\approx \frac{1}{8\,\pi^{2}}
\!\!\!\!\!\!\!\int\displaylimits_{Supp(P_{1})}\!\!\!\!\!\!\!\! d\kappa^{2}\,
\sqrt{\kappa^{2}}\e^{2\mu_{\kappa}\tau}
\ .
\end{equation}
The Floquet index as a function of $\kappa^2$ has a parabolic shape
around $\kappa_{1}$, 
\begin{equation}
\mu_\kappa^{(1)}\approx \mu_{\kappa_1}^{(1)}\left[
1-\left(\frac{\kappa^{2}-\kappa_{1}^{2}}{\Delta\kappa_1^2/2}\right)^2\right]
\ ,
\end{equation}
where $\Delta\kappa_1^2$ denotes the width of the first peak (the same notation we used for
the fermions) which scales with the Higgs vacuum total energy exactly according to the
same law \eqref{DKscale} for fermions provided $c\lesssim -1+10^{-3}$.
\par
For $\mu_{\kappa}\tau\ll 1$, the integral in Eq.~\eqref{BON_approx_s} can now be easily
estimated and yields
\begin{equation}
\label{BON_s}
n_{B}(\tau)
\approx
\frac{\left(\mu_{\kappa_1}^{(1)}\right)^2}{2\,\pi^2}\,N_B\,\tau^2
\ ,
\end{equation}
where $N_B=N_B(\kappa_1,\Delta\kappa_1^2)$ is a rather cumbersome expression
which we do not show explicitly since it will not be used in the present paper.
For long times, one can likewise estimate the integral in Eq.~\eqref{BON_approx_l} using
a \emph{saddle point} approximation and obtain \cite{kof-linde-starob}
\begin{equation}
\label{BON_l}
n_{B}(\tau)
\approx
\frac{\kappa_{1}\,\Delta\kappa_1^2}{16\,\pi^{2}}
\sqrt{\frac{\pi}{\mu_{\kappa_1}^{(1)}}}\,
\frac{\e^{2\,\mu_{\kappa_1}^{(1)}\tau}}{\sqrt{2\,\tau}}
\ .
\end{equation}
\par
%
%
%
\begin{figure}[t]
   \centering
   \subfigure
   {\includegraphics[scale=0.8]{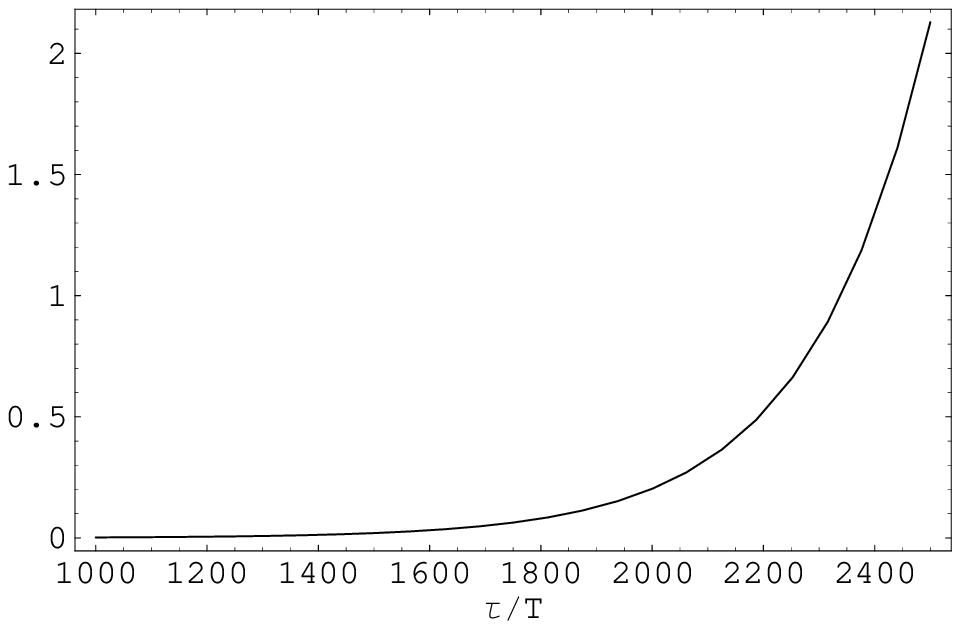}}
    \subfigure
   {\includegraphics[scale=0.8]{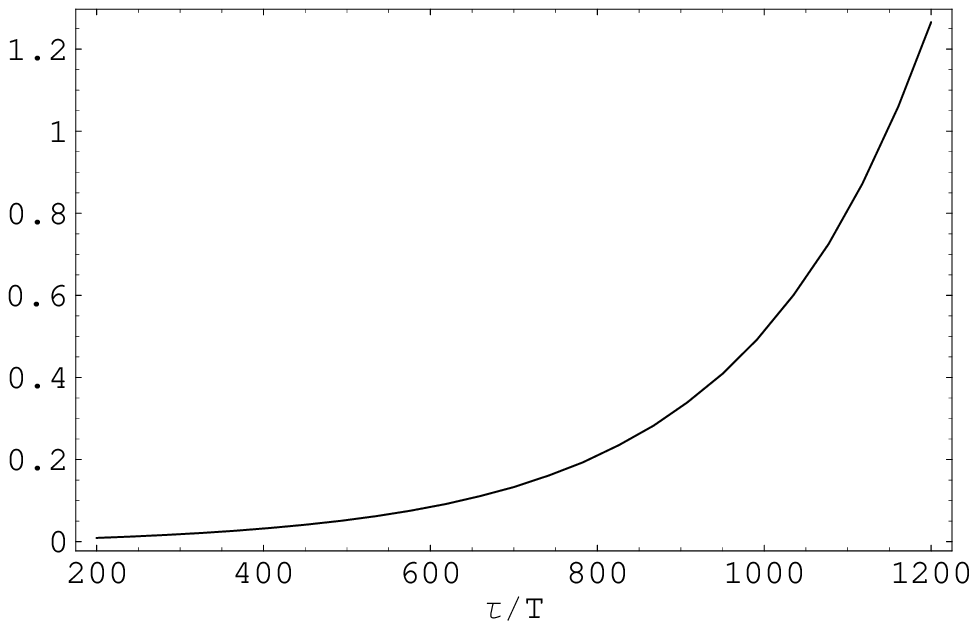}}
 \caption{Time evolution of the total number of produced $Z^0$ (left graph)
 and fraction of the Higgs vacuum energy absorbed by $Z^0$
 (right graph) for $M_H=500\,$GeV and $c_0=-1+10^{-3}$
 (in the right plot, fractions larger than unity signal that the back-reaction
 cannot be neglected).}
\label{plot_Z0}
\end{figure}
The left plot in Fig.~\ref{plot_Z0} shows the time dependence of the total number
of produced $Z^0$ following from \eqref{BON_l} for $M_H=500\,$GeV and $c_0=-1+10^{-3}$.
This quantity grows exponentially in time and when the fraction~\footnote{This fraction is defined
in analogy with that for the fermions given in Eq.~\eqref{R}.} of Higgs vacuum energy
converted into $Z^0$ approaches unity (see right plot), the back-reaction is expected
to become relevant.
Indeed, values of that fraction larger than unity do not make sense and actually
signal the complete failure of the approximation which neglects the back-reaction.
Note that this occurs after about a thousand background oscillations.
Recalling the results for the {\em top} from Section~\ref{fermi_pro},
it is clear that, during the first few hundreds of vacuum oscillations
the energy transferred to the \emph{top} is larger than that absorbed by
the $Z^0$, and it seems that the back-reaction of the latter particles can be neglected at an early stage.
The production of $Z^0$ and their interaction with the time dependent Higgs
background take the lead later and, with a good approximation, remain the only processes
with significant effects (we shall have more to say on this in the next Section).
%
\section{Fermion and boson back-reaction}
\label{back}
\setcounter{equation}{0}
The results presented so far have been obtained by neglecting the back-reaction of
the produced fermions and bosons on the evolution of the Higgs vacuum.
This is a common approximation, since a more complete treatment of these
effects is a very difficult task.
We shall therefore begin by considering the simple case of one kind of fermion
simply denoted by $\psi$ (which we shall identify with the \emph{top} quark later)
and one kind of boson (the $Z^0$). 
\par
The Lagrangian density for our system is given by  (see Eqs.~\eqref{lagr_phi,h}, 
\eqref{lagr_fermioni} and~\eqref{lagrWZ} with $v = M_{H}\Phi/{\sqrt{\lambda}}$)
\begin{equation}
\label{L_BRfer}
\mathcal{L}_{BR}=
{\mathcal L}_{H}
 -\frac{\lambda}{4}\,\sqrt{q}\,\bar{\psi}{\psi}\,\Phi
      -\frac{\lambda}{8}\,q\,Z^{\mu}Z_{\mu}\,\Phi^2
      \ ,
\end{equation}
where we have redefined the product
$\langle\bar{\psi}{\psi}\rangle$ to make it dimensionless.
The back-reaction effects can be studied to a good degree of accuracy in the
Hartree approximation (see Refs.~\cite{kof-linde-starob,Boyan-deVega}
for a discussion on this subject).
Let us then consider the vacuum expectation value of the
Euler-Lagrange equation for $\Phi$ which reads
\begin{equation}
\label{eq_phiBR}
 \Phi''-2\Phi(1-\Phi^{2})+
  \frac{\lambda}{4}\,\sqrt{q}
         \langle\bar{\psi}{\psi}\rangle
         -\frac{\lambda}{4}\,q\,\langle Z^{\mu}Z_{\mu}\rangle \Phi
         =0
         \ .
\end{equation}

\par
The product $\langle\bar{\psi}{\psi}\rangle$ can be rewritten in terms of an
integral in momentum space of the Bogoliubov coefficients (previously
introduced to diagonalize the Hamiltonian operator for a generic fermion
field) using the expansion~\eqref{sviluppo_psi}.
In this way one encounters ultraviolet divergencies and, in order to obtain a finite
result, the operator $\bar{\psi}{\psi}$ must be normal ordered to subtract 
vacuum quantum fluctuations~\cite{Boyan-deVega,kof-linde-starob,giudice-riotto}.
Moreover, we shall neglect any other renormalization related, for example,
to perturbative quantum corrections.
Note that after a Bogoliubov transformation the vacuum state
$|0_\tau\rangle$ is time-dependent and we have a different renormalization
at every time.
We thus define the (time-dependent) normal ordering of a generic operator as
\begin{equation}
\mathscr{N}_{\tau}(\mathcal{O})\equiv \mathcal{O}
                  -\langle0_{\tau}|\mathcal{O}|0_{\tau}\rangle,
\end{equation}
so that the \emph{vev} of $\bar{\psi}{\psi}$ will be given by
\begin{eqnarray}
\label{termineBR}
 \langle \bar{\psi}\psi \rangle \equiv
            \langle0|\mathscr{N}_{\tau}(\bar{\psi}\psi)|0\rangle 
   &=&\langle0|\bar{\psi} \psi |0\rangle
      - \langle0|\left(\langle0_{\tau}|\bar{\psi} \psi |0_{\tau}\rangle\right)|0\rangle
       \nonumber \\ 
   &=&\langle0|\bar{\psi} \psi |0\rangle 
       -\langle0_{\tau}|\bar{\psi} \psi |0_{\tau}\rangle
       \ .
\end{eqnarray}
Using results from Section~\ref{sec-fermion_theory}, we find
\begin{eqnarray}
\label{fermiBRterm}
  \langle \bar{\psi}\psi \rangle 
    &=&\int\frac{d^{3}\kappa}{2\,\pi^{3}}\, 
    \kappa^{2}\left[\left|X_{\kappa}(\tau)\right|^{2}
      +\frac{\sqrt{q}\,\Phi(\tau)}{2\,\omega_{\kappa}(\tau)}-\frac{1}{2}
    \right]
     \nonumber
     \\
    &=&-2\,n_{\psi}(\tau) + \int \frac{d^{3}\kappa}{2\,\pi^{3}}\,
   \kappa^{2}\left[\left|X_{\kappa}(\tau)\right|^{2}
     -\frac{\im\left[ X_{\kappa}(\tau)X_{\kappa}^{'*}(\tau) \right]}{2\,\omega_{\kappa}(\tau)}
   \right].
\end{eqnarray}
As for the $Z^0$, analogous prescriptions to those used for the fermion lead to
\begin{equation}
\label{bosonBRterm}
\langle Z^{\mu}Z_{\mu}\rangle
=-\int\frac{d^3\kappa}{(2\,\pi)^{3}}  
      \left(\left|Y_{\kappa}\right|^{2}-
      \frac{1}{\sqrt{\kappa^{2}+q\,\Phi^{2}}}\right)    
\ .
\end{equation}
Note that the two back-reaction terms~\eqref{fermiBRterm} and
\eqref{bosonBRterm} vanish at $\tau=0$ by virtue of the initial
conditions~\eqref{condizioni_iniziali_adim}
and~\eqref{cond_iniz_bos},
as expected.
The system of back-reaction equations is thus
\begin{equation}
\label{sistemaBR}
\left\{
\begin{array}{l}
 \Phi''-2\Phi(1-\Phi^{2})
   +\strut\displaystyle\frac{\lambda}{4}\sqrt{q}
        \langle \bar{\psi}\psi \rangle
-\frac{\lambda}{4}\,q\,\langle Z^{\mu}Z_{\mu}\rangle \Phi
 =0
        \\
        \\
  X''_{\kappa}
  +\left( \kappa^{2}+q\,\Phi^{2}-i\sqrt{q}\,\Phi'
   \right)X_{\kappa}=0
   \\
   \\
  Y''_{\kappa}
  +\left( \kappa^{2}+q\,\Phi^{2}
   \right)Y_{\kappa}=0
   \ . 
\end{array} 
\right.
\end{equation}
The last two terms in the first equation (Higgs-fermion and Higgs-vector coupling terms)
depend on $\lambda$, that is the strengh of the quartic term in the Higgs \emph{vev}
potential.
In the Standard Model the coupling $\lambda$ is not fixed but
it is possible to take $0<\lambda < 1$ if one considers the Higgs self-interaction
as described by a perturbative theory.
\par
For the above system, it is easy to show that the (renormalized) total energy 
\begin{eqnarray}
\label{EBR}
E=
{M_H}\left\{
\frac{c}{\lambda}+
\frac{1}{2}\int \frac{d^3\kappa}{(2\,\pi)^3}\,\omega_\kappa\,n_\kappa^{(f)}
+\frac{1}{2}\int \frac{d^3\kappa}{(2\,\pi)^3}\,\varpi_\kappa\,n_\kappa^{(B)}
\right\}
\ ,
\end{eqnarray}
is exactly conserved by virtue of the equations of motion themselves.
In the above
\begin{equation}
\label{BRvacuumenergy}
  c(\tau)=\left[\Phi'(\tau)\right]^{2}-2\,\Phi^{2}(\tau)+\Phi^{4}(\tau)
  \ ,
\end{equation}
the fermion (boson) occupation number $n_\kappa^{(f)}$ ($n_\kappa^{(B)}$)
and frequency $\omega_\kappa$ ($\varpi_\kappa$) are given in
Eq.~\eqref{numero_occupazione_adim} [Eq.~\eqref{no_bos_esatto}] and
Eq.~\eqref{om_f} [Eq.~\eqref{om_b}], respectively.
\par
The approximate evaluation of the integrals~\eqref{fermiBRterm} and~\eqref{bosonBRterm}
by restricting their integrand on the dominant band (see also ~\cite{kof-linde-starob}) is
based on the results of Sections~\ref{fermi_pro} and \ref{boso_pro} for fermions
and bosons respectively, and proceeds in a similar way.
We therefore assume the validity of scaling laws of the form given, for example,
in Eqs.~(\ref{DKscale})
or (\ref{n(energy)}) and Eqs.~(\ref{BON_s}) and (\ref{BON_l}).
These approximations are not valid for asymptotically long time evolution, when
the dissipative dynamics takes over and very little particle production may still
take place.
In such a case, one may need different computational schemes, and probably a full lattice
approach, useful also to describe rescattering phenomena and estimate a possible
termalization phase.
Nevertheless, our approximation scheme has the virtue of been simple enough
to lead to a system of differential equations of finite order which helps to grasp some
aspects of the back-reaction dynamics and gains more and more validity in the limit
$\lambda \ll 1$. 
\par
In order to make the qualitative features of the system~\eqref{sistemaBR} clearer,
we shall first study separately its behaviour for short and long times.
In particular, as we have seen in the two previous Sections, 
the production of fermions is expected to dominate during the first few hundreds
of Higgs background oscillations and in that regime we shall neglect
boson production by simply switching its coupling off and set $M_H=500\,$GeV
in order to have the maximum \emph{top} production.
At larger times the production of bosons overcomes that of fermions
(limited by Pauli blocking) and we shall then neglect the fermions
and set $M_H=200\,$GeV so as to maximise the $Z^0$ production.
In the last Subsection, we shall finally consider the entire system
with both the \emph{top} and $Z^0$ for $M_H=500\,$GeV in order to show
the effect of the fermion back-reaction on the later boson production.
This value of the Higgs mass, although not very likely, is chosen as it is
particularly convenient to study the interplay between fermion and boson production. 
\begin{figure}[t]
   \centering
   \subfigure
   {\includegraphics[scale=0.8]{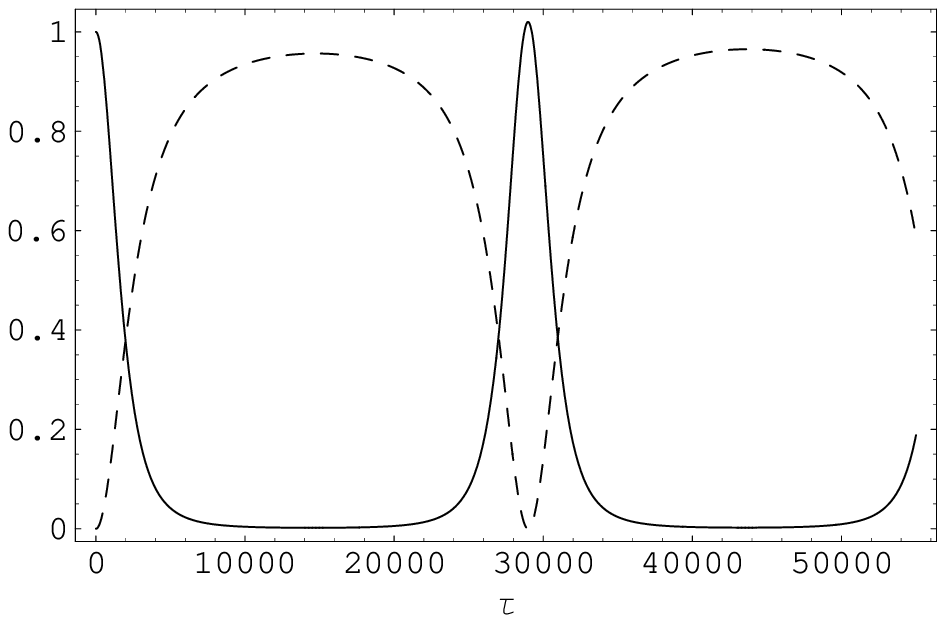}}
   \subfigure
   {\includegraphics[scale=0.8]{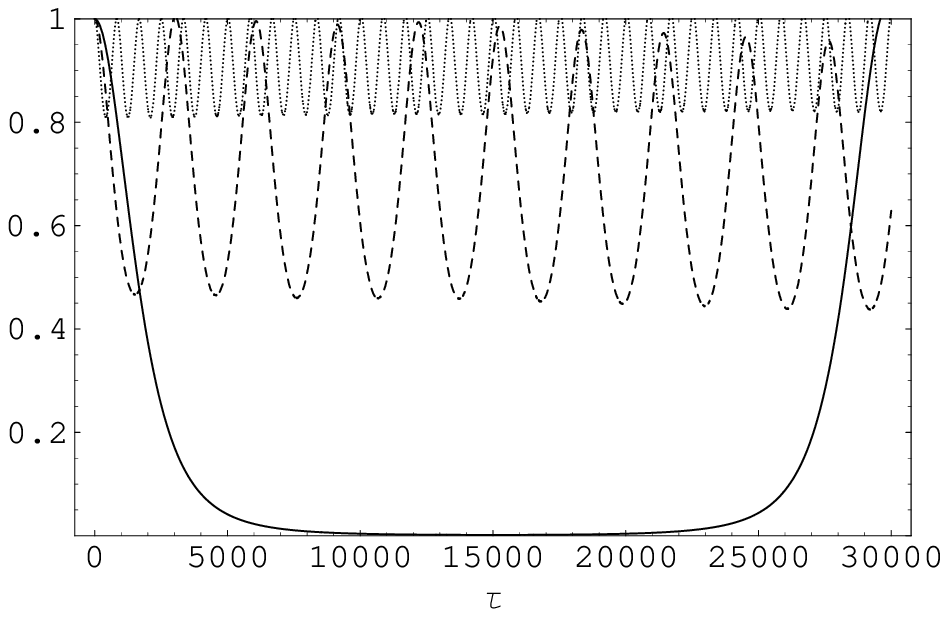}}
   \caption{{\it Fermion back-reaction} with $M_H=500\,$GeV.
   {Left plot}: Time evolution of the {\em top} occupation number (dashed line)
   and Higgs vacuum energy (solid line) for $\lambda=1$ and initial $c(0)=-1+10^{-5}$.
   {Right plot}:
   Comparison of the Higgs total vacuum energy for initial values  $c(0)+1=10^{-3}$
   (dotted line), $10^{-4}$ (dashed line) and $10^{-5}$ (solid line).
   A normalization factor of $[c(0)+1]^{-1}$ is used for the Higgs energies.}
\label{energyBR}
\vskip 0.5cm
%
   \centering
   {\includegraphics[scale=0.80]{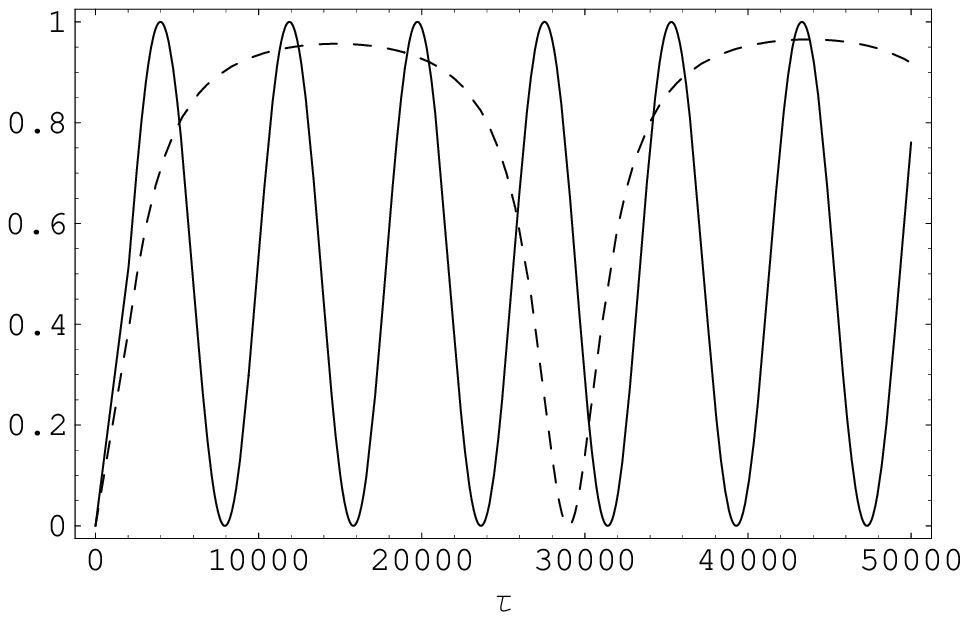}}
   \caption{{\it Fermion back-reaction}:
            comparison of {\em top} occupation number with (dashed line) and
            without (solid line) back-reation for $c(0)=-1+10^{-5}$.}
\label{compF}
\end{figure}
%
\subsection{Fermion back-reaction}
\label{fermi_back}
As we mentioned above, for relatively short times, we at first consider the evolution
of the Higgs background and one fermion, namely the \emph{top} quark with
$M_{H}=500\,$GeV.
\par
The left graph in Fig.~\ref{energyBR} shows the time dependence of the corresponding
occupation number for the \emph{top} and Higgs energy densities up to a time
$\tau\simeq 3.14\cdot 10^{3}$ ($\simeq 10^3\,T$, where $T$ is the Higgs period without
back-reaction)
when the boson back-reaction is expected to take over.
The Higgs energy oscillates periodically in time and of course takes its minimum
values when the {\em top}'s occupation number is maximum.
The right graph displays how the Higgs energy changes in time
for different initial values of $c$, that is for different total energy
(all curves assume $\lambda=1$).
Note that the maximum fraction of Higgs energy converted into the \emph{top}
increases for decreasing initial total energy, as one would expect from the Pauli
blocking.
Fig.~\ref{compF} then compares the \emph{top}'s occupation number
with and without back-reaction for $c(0)=-1+10^{-5}$.
It is clear that the back-reaction in general suppresses the number of produced
fermions and this effect is more pronounced for smaller total energy.
We also remark that the above numerical solution conserves the total energy
with an accuracy better than 1 part in $10^3$.
\par
To conclude this section, we would like to spend a few words about the
behaviour of the system with respect to the value of $\lambda$.   
Further numerical analysis shows that the fraction of energy absorbed by the
produced fermions is proportional to $\lambda$. 
Moreover, the system of non-linear differential
equations~\eqref{sistemaBR} seems to produce a chaotic
behaviour only for $\lambda>10^{6}$.
In fact for $\lambda \approx 10^{6}$ the highly non-linear term proportional
to $\langle\bar{\psi}\psi\rangle$ becomes of the same
order of magnitude as the other terms in the first equation
of~\eqref{sistemaBR}.
But at this level the approximations used in our equations to evaluate the
back-reaction terms should break down. 
%
%
\subsection{Boson back-reaction}
\label{boso_back}
From the results of the previous Section, we know that the energy density of
created bosons is comparable to the initial Higgs energy density after
about $10^{3}$ vacuum oscillations (that is, for $\tau\gtrsim 3.14\cdot 10^{3}$).
The production of bosons then overcomes fermion production (limited by the Pauli
principle) and their back-reaction becomes the most relevant phenomenon.
We shall therefore neglect the fermion contribution here and just consider the $Z^0$
and $M_H=200\,$GeV or $M_H=500\,$GeV.
\par
Even with the above simplification, it is still very difficult to solve 
the system~\eqref{sistemaBR} and we need to employ yet another approximation.
Since the integrand in Eq.~\eqref{bosonBRterm} is sharply peaked around the centres
of resonance bands, we estimate that integral in the same
way we used to obtain the total number of produced bosons in
Eq.~\eqref{BON_l}, as we have already anticipated in the general discussion of
this Section.
\par
%
\begin{figure}[ht!]
   \centering
   \subfigure
   {\includegraphics[scale=0.80]{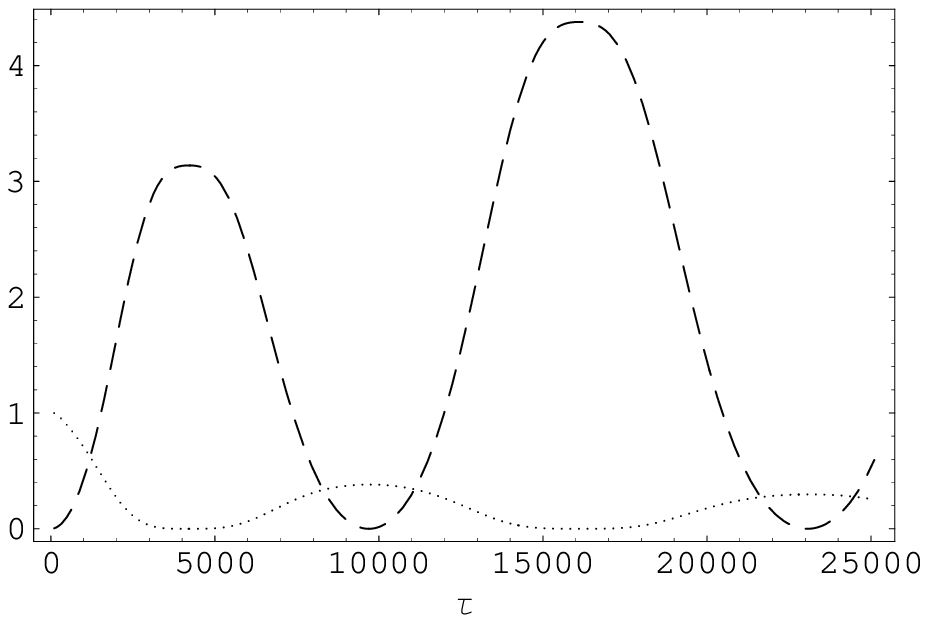}} 
   \subfigure
   {\includegraphics[scale=0.8]{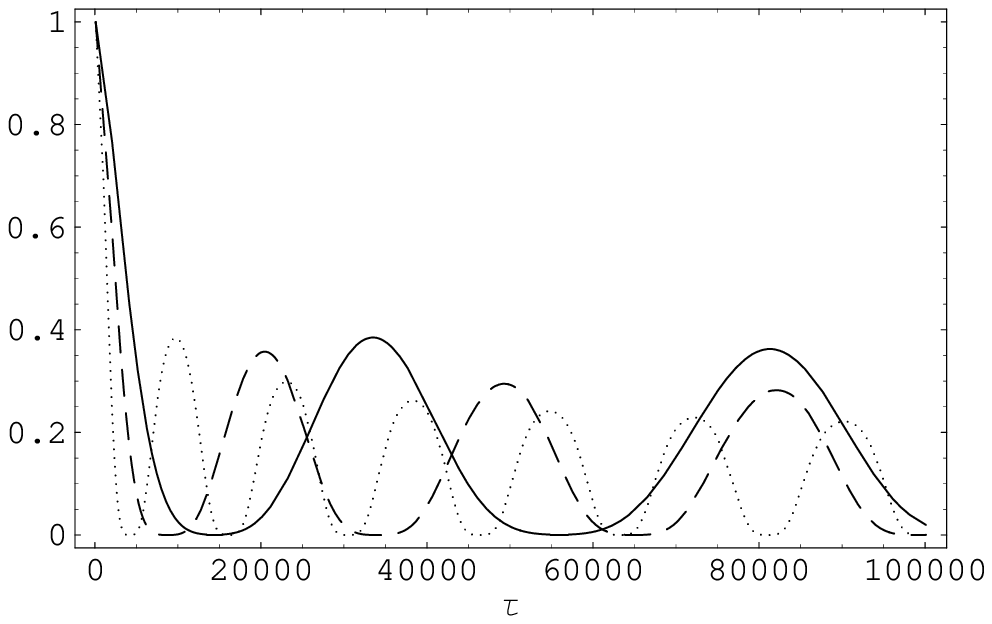}}
   \caption{{\it Boson back-reaction} with $M_H=200\,$GeV.
   {Left graph}: $Z^0$ occupation number (dashed line) and Higgs vacuum energy
   (dotted line) for an initial value of $c(0)+1=10^{-5}$.
   {Right graph}:
   Time variation of the Higgs vacuum energy
   for initial total energy $c(0)+1=10^{-5}$ (dotted line),
   $10^{-6}$ (dashed line) and
   $10^{-7}$ (solid line).
   A normalization factor of $[c(0)+1]^{-1}$ for the Higgs energies is always used.
}
\par
\label{energyBRboson}
   \centering
\subfigure
   {\includegraphics[scale=0.80]{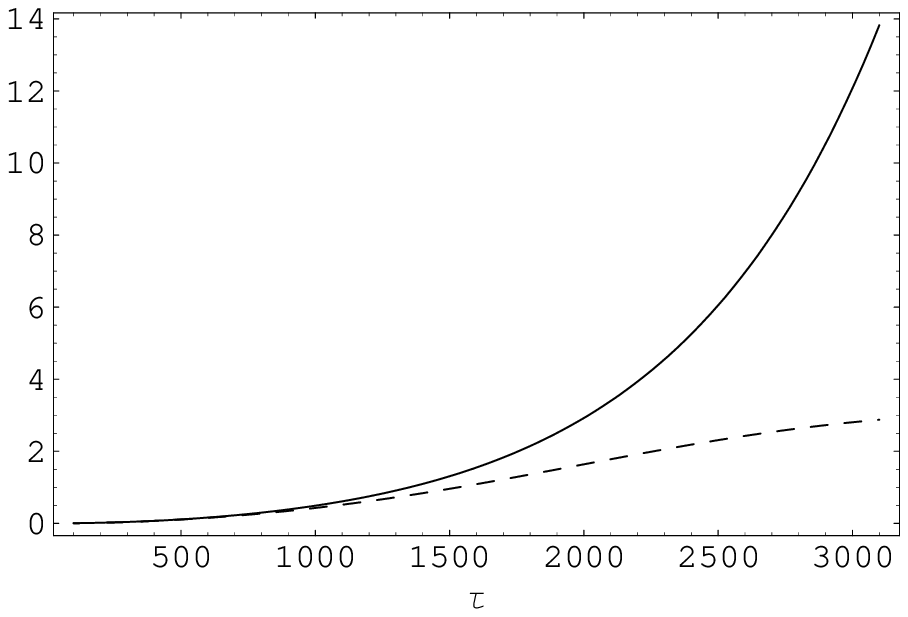}}
\subfigure
   {\includegraphics[scale=0.80]{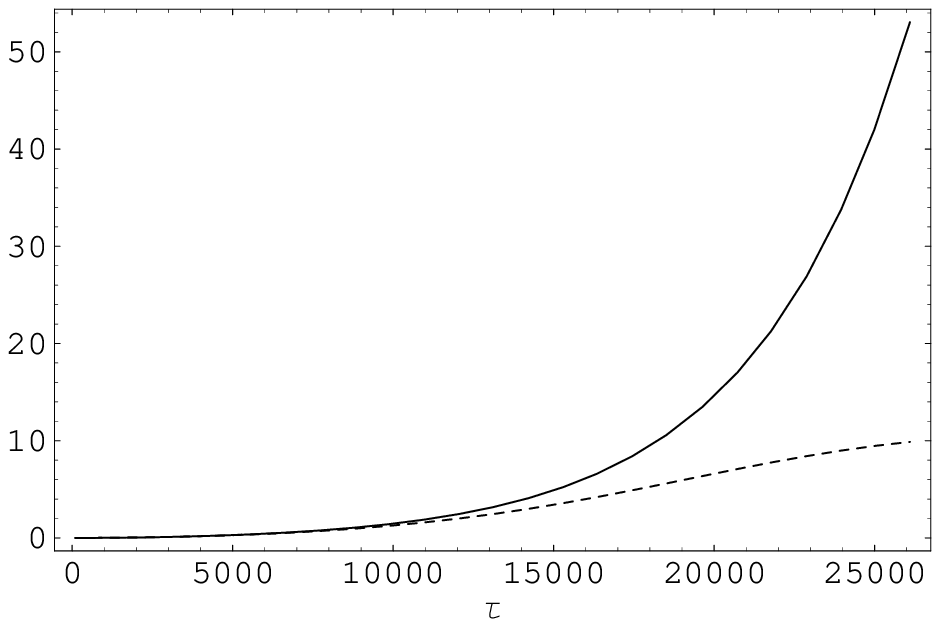}}
   \caption{{\it Boson back-reaction}: comparison of $Z^0$ occupation numbers
   with (dashed line) and without (solid line) back-reation for $c(0)=-1+10^{-5}$.
   $M_H=200\,$GeV in left plot and $M_H=500\,$GeV in right plot.}
\label{compB}
\end{figure}
We can now solve numerically the system~\eqref{sistemaBR}
and Fig.~\ref{energyBRboson} shows some relevant results.
Starting from the left, it is possible to see that
the Higgs energy is minumum when the rate of production is maximum. 
Further, from Fig.~\ref{compB}, one sees that the back-reaction strongly suppresses
the boson production.
The right plot in Fig.~\ref{energyBRboson} shows the amount of vacuum energy
dissipated by back-reaction effects as a function of the time for three
different values of the total (dimensionless) energy (which of course coincides with
the initial Higgs energy).
An important feature of this plot is that we have a ``regeneration'' of the
vacuum energy, which now oscillates in time,  and the number of peaks in a given time
interval is seen to increase (that is, the periods of oscillations become shorter)
for increasing total energy.
In all cases, the values at the peaks slowly decrease in time and the periods of
oscillations stretch.
One can therefore conclude that the system will evolve towards a complete
dissipation of the Higgs oscillations. 
This is due to the presence of a dumping term [analogous to $1/\sqrt{\tau}$
in Eq.~\eqref{BON_l}] that appears in the back-reaction term~\eqref{bosonBRterm}
after integrating in $\kappa$.
%
\subsection{Fermion and boson back-reaction}
\label{tot_back}
\begin{figure}[hb!]
   \centering
   \subfigure
   {\includegraphics[scale=0.80]{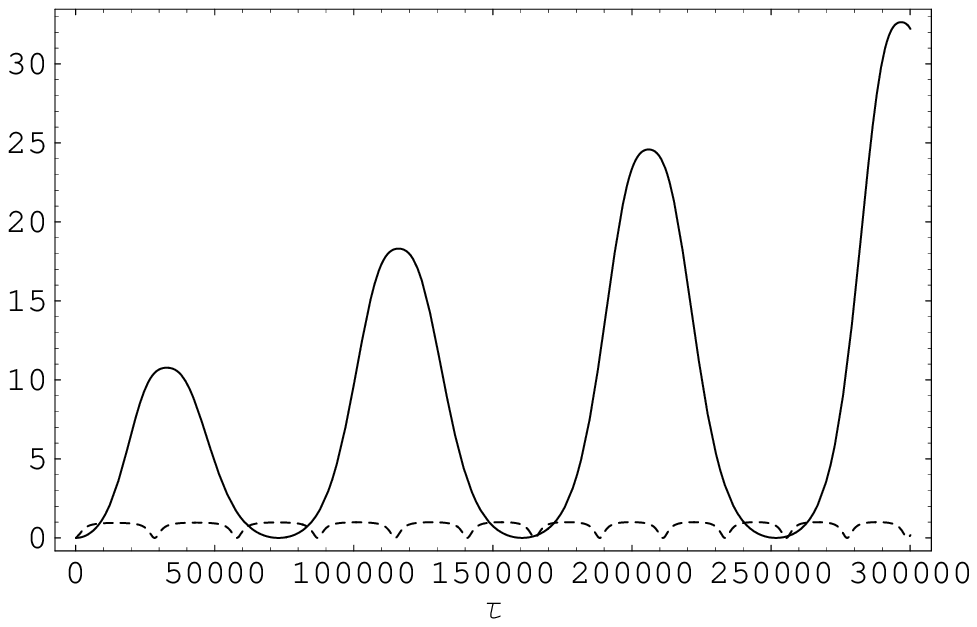}}
   \subfigure
   {\includegraphics[scale=0.80]{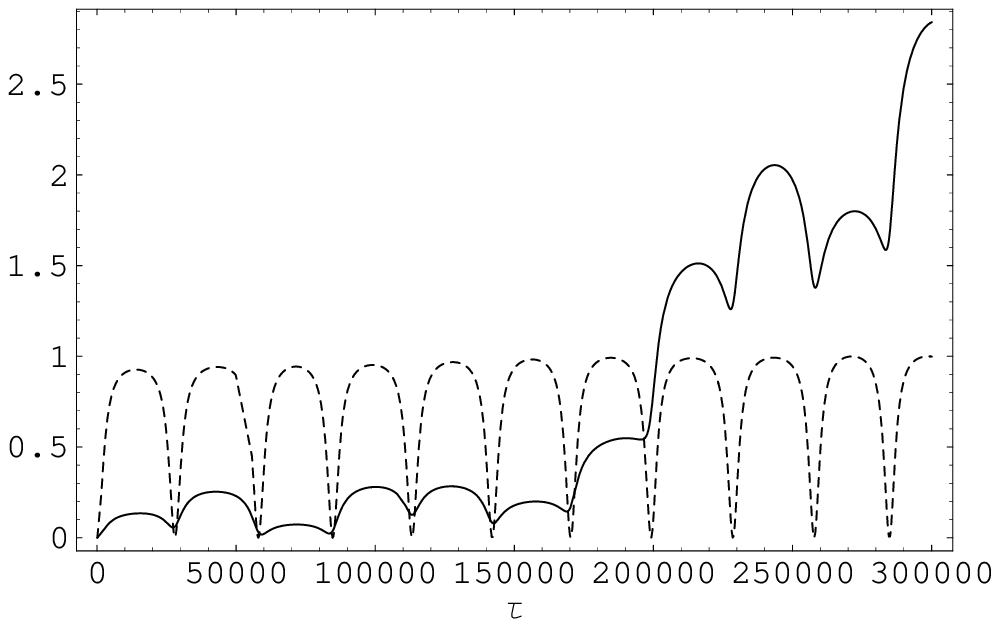}}
   \caption{Time evolution of the occupation numbers for the {\em top}
   (dashed line) and $Z^0$ (solid line) for $\lambda=1$, $M_H=500\,$GeV and
   $c(0)=-1+10^{-5}$. Left plot: {\it Partial back-reaction} from
   the approximations of Sections~\ref{fermi_back} and \ref{boso_back}.
    Right plot: {\it Full back-reaction} from Section~\ref{tot_back}.}
\label{BR_c}
\end{figure}
%
We finally consider the whole system with the \emph{top} and $Z^0$ described by
Eq.~\eqref{sistemaBR} for $M_H=500\,$GeV.
Again, we choose this value of the Higgs mass in order to have both fermion and boson
production from the first resonant band.
This corresponds to a particularly complicated scenario with large production of
both the \emph{top} and $Z^0$, whereas a more realistic smaller value of $M_H$
would mostly lead to the production of lighter quarks with smaller effects on the
$Z^0$ at late times.
\par
The coupled dynamics is then described in Fig.~\ref{BR_c} in which we show
on the right the occupation numbers for the \emph{top} and $Z^0$.
This results has to be compared to the results of
back-reaction computed in the ``factorized'' approximation used in Secs. $7.1$
and $7.2$ which we report together for convenience on the left,
both computed at $M_H=500\,$GeV.
We note that the \emph{top} production is not particularly affected,
whereas the $Z^0$ occupation number is suppressed in the sense that
it increases significantly only at later times.
The time needed for the production of $Z^0$ to overtake the \emph{top}
is about one order of magnitude larger than without fermion back-reaction.
\par
As we have already observed, we are able to study the evolution of the system
in a window of time far from the asymptotics (presumably with an upper bound
in time inversely proportional to $\lambda$).
Beyond that, the adopted approximations break down and the asymptotic time
behaviour of the system should be studied taking into account a more accurate
description of the mode production (eventually in a full lattice approach) together
with the rescattering contribution of the produced particles. 
%
\section{Conclusions}
\label{conc}
\setcounter{equation}{0}
We have considered oscillating solutions for the Higgs \emph{vev}
in the context of the Standard Model of particle physics and studied some
resulting dissipative effects.
In the Standard Model, in fact, the masses of fundamental particles depend on 
that \emph{vev} and its oscillations can be viewed as a time-dependent
renormalization of the particles' masses which leads to the production of
fermion and boson pairs by parametric resonance.
\par
In the first part of the present paper, the back-reaction of the produced
pairs has been totally neglected.
In this approximation, particle production by the oscillating Higgs appeared
to be very efficient.
For fermions, the Pauli blocking constrains their occupation numbers
to oscillate in time (see Figs.~\ref{fig:NO&detNO}) about mean values smaller
than one.
From the entries given in Table~\ref{tabelle_prob_di_prod}, one can see that
such mean values strongly depend on the resonance parameter $q$,
that is the Higgs and fermion masses [see the definitions~\eqref{parametri_adim}].
Moreover, the particular form of the governing Dirac equation~\eqref{eq_modi}
yields a significant probability of producing only those fermions whose momenta
lie on well-defined bands in the $(q,\kappa^2)$-plane
[see Fig.~\ref{cartainst} and Eq.~\eqref{k_n_fermi}].
Considering the masses of Standard Model particles, this also implies
a larger probability of producing fermions with non-relativistc momenta.
\par
As for the bosons, their production is not constrained by any fundamental 
principles and the governing Mathieu equation~\eqref{mathieu} (in the small
oscillation regime) leads to occupation numbers which grow exponentially in time.
Analogously to the fermions, the boson production only occurs on narrow bands in the
$(q,\kappa^2)$-plane [see the definitions~\eqref{boso_par} and Eq.~\eqref{k+q=n}].
\par
In our analysis, we have regarded the Higgs mass $M_H$ as an adjustable
parameter, for the experimental data only place a lower bound on it.
Consequently, we have given our results for different possible values of
$M_H$ and the initial Higgs energy $c(0)$ [see the definition~\eqref{phi_eq_1}
and Eq.~\eqref{BRvacuumenergy}].
For natural values of these parameters, we have found that a significant
fraction of the initial vacuum energy can be transferred to the fermions
before a complete background oscillation (whose period we denoted as $T$,
see Fig.~\ref{plot_densEnFer}).
The Pauli blocking then prevents the fermion production from increasing
further.
The bosons, on the other hand, grow exponentially in time but absorb
a significant fraction of the Higgs vacuum energy only after about
a thousand oscillations (see Fig.~\ref{plot_Z0}).
This feature seems to suggest that it is possible to analyse the
back-reaction of fermions and bosons separately.
\par
In particular, we have studied the system of one fermion (the \emph{top} quark)
and one boson (the $Z^0$) coupled to the Higgs field $\Phi$
in the Hartree approximation [see Eq.~\eqref{sistemaBR}].
For such a system, the total energy~\eqref{EBR} is conserved and,
for $0\le\tau\lesssim 10^3\,T$,
the back-reaction of the produced $Z^0$ has been neglected.
To make the problem more tractable, and avoid a complicated study of the system
on a lattice of Fourier modes, we have adopted another approximation which
might breakdown at asymptotic times. 
We have thus found that the fermion production is in general mildly suppressed
by the back-reaction (see Figs.~\ref{energyBR} and~\ref{compF}).
At later times ($\tau\gtrsim 10^3\,T$), the back-reaction of the produced
\emph{top} has been neglected and, in the same framework of approximations,
we have found that the production of $Z^0$ is more strongly suppressed by the
back-reaction (see Figs.~\ref{energyBRboson} and~\ref{compB}).
\par
The study of the system where both fermion and boson backreaction effects are
included has shown a slower boson
production, which takes over almost an order of magnitude later in time,
against the naive expectation of some kind of factorization in the production
events.
We however remark that we have chosen a value of the Higgs mass $M_H=500\,$GeV
which corresponds to a large production of the \emph{top} in order to study
its effect on the production of the $Z^0$.
For a more realistic (presumably smaller) value of $M_H$, one should instead 
consider the quarks and leptons with the higher production rates (as shown in
Table~\ref{tabelle_prob_di_prod}).   
\par
Future investigations, apart from being focused on more appropriate values of
the Higgs mass as, for example, an awaited discovery at LHC would provide,
should address aspects which we have found beyond our possibility because
of the approximation schemes adopted.
In particular, one should investigate the production and dissipative dynamics in
the late (asymptotic) times of the evolution and evaluate the rescattering phenomena
associated with the produced particles.
Keeping in mind that we are mainly considering a possible ``starting'' regime of very
small Higgs oscillations, the rescattering could lead to a thermal
background characterized by a very low temperature.
\par
We conclude by noting that, although suppressed when the back-reaction is
properly included in the analysis, particle production by parametric resonance
with an oscillating Higgs field remains a remarkable effect with phenomenological
relevance and could be an (indirect) way of testing the time-dependence of the
constants of the Standard Model.
%
%
%

%

\begin{thebibliography}{99}
\bibitem{uzan}
J.-P.~Uzan, Rev. Mod. Phys. \textbf{75}, 403 (2003).
%
\bibitem{dirac-old}
P.~A.~M.~Dirac, Nature (London), \textbf{139}, 323 (1937);
Proc. Roy. Soc. London \textbf{A165}, 198 (1938). 
%
\bibitem{dirac-new}
P.~A.~M.~Dirac, Proc.~Roy.~Soc.~London \textbf{A338}, 439 (1974);
Proc. Roy. Soc.~London \textbf{A365}, 19 (1979).
%
\bibitem{kujat}
J.~Kujat, R.~J.~Scherrer, Phys.~Rev.~D \textbf{62}, 023510 (2000).
%
\bibitem{hannestad}
S.~Hannestad, Phys.~Rev.~D \textbf{60}, 023515 (1999).   
%
\bibitem{kaplinghat}
M.~Kaplinghat, R.~J.~Scherrer e M.~S.~Turner,
Phys.~Rev.~D \textbf{60}, 023516 (1999). 
%
\bibitem{Taylor-Veneziano}
T.~R.~Taylor and G.~Veneziano, Phys.~Lett. \textbf{B 213}, 450 (1988). 
%
\bibitem{Witten}
E.~Witten, Phys.~Lett. \textbf{B 149}, 351 (1984). 
%
\bibitem{passarino}
G.~Passarino,
``Are constants constant?,'' [hep-ph/0108254].
%
\bibitem{prod}
V.~M.~Mostepanenko, V.~M.~Frolov and V.~A.~Shelyuto,
Yad.\ Fiz.\  {\bf 37} (1983) 1261;
A.~D.~Dolgov and D.~P.~Kirilova,0
Sov.\ J.\ Nucl.\ Phys.\  {\bf 51} (1990) 172
[Yad.\ Fiz.\  {\bf 51} (1990) 273].
%
\bibitem{higgs-mass}
F.~Cerutti, in \emph{Proc.~of the 16th Rencontres de Physique
de la Vallee d'Aoste: Results and Perspectives in Particle Physics},
La Thuile, Valle d'Aosta, Italia, 3 -- 9 Marzo 2002;
CERN - ALEPH - PUB - 2002 - 003. [hep--ex/0205095];
\\
U.~Baur et al., \emph{Summary Report of the Precision Measurement
Working Group at Snowmass} 2001 [hep--ph/0202001];
\\
Yu.~F.~Pirogov, O.~V.~Zenin, Eur.~Phys.~J. \textbf{C10}, 629 (1999);
\\
U.~Mahanta,
``New physics, precision electroweak data and an upper bound on Higgs
mass,''
[hep-ph/0009096].
%
\bibitem{ba}
J.~Baacke, K.~Heitmann and C.~Patzold,
Phys.\ Rev.\  D {\bf 58} (1998) 125013.
%
\bibitem{Mostepanenko_book}
V.~M.~Mostepanenko, A.~A.~Grib, V.~M.~Mamayev,
\emph{Vacuum Quantum Effects in Strong Fields}, Friedmann Laboratory Publishing,
San Pietroburgo (1994).
%
\bibitem{Boyan-deVega}
D.~Boyanovsky, H.~J.~de Vega, R.~Holman, D.~S.~Lee and A.~Singh,
Phys.\ Rev.\ D {\bf 51} (1995) 4419.
%
\bibitem{kof-linde-starob}
L.~Kofman, A.~Linde and A.~A.~Starobinsky,
Phys.~Rev.~D \textbf{56}, 3258 (1997).
%
\bibitem{heat}
J.~Garcia-Bellido, S.~Mollerach and E.~Roulet,
JHEP {\bf 0002} (2000) 034;
P.~B.~Greene, L.~Kofman, A.~Linde and A.~A.~Starobinsky,
Phys.~Rev.~D \textbf{56}, 6175 (1997);
P.~B.~Greene and L.~Kofman,
Phys.\ Lett.\  B {\bf 448} (1999) 6;
D.~Boyanovsky, H.~J.~de Vega, R.~Holman and J.~F.~J.~Salgado,
Phys.\ Rev.\  D {\bf 54} (1996) 7570.
%
\bibitem{boya2}
  D.~Boyanovsky, M.~D'Attanasio, H.~J.~de Vega, R.~Holman and D.~S.~Lee,
  Phys.\ Rev.\ D {\bf 52} (1995) 6805.
%
\bibitem{SMpassarino}
D.~Bardin and C.~Passarino, \emph{The Standard Model in
the making: Precision study of the electroweak interaction}, Oxford UK,
Clarendon (1999).
%
\bibitem{phi4}
J.~A.~Espichan Carrillo, A.~J.~Maia and V.~M.~Mostepanenko,
Int.\ J.\ Mod.\ Phys.\  A {\bf 15} (2000) 2645.
%
\bibitem{Gradsh_Ryz}
I.~S.~Gradshteyn, I.~H.~Ryzhik,
\emph{Table of Integrals, Series and Products}, Academic Press, New York
(1980).
E.~T.~Whittaker, G.~N.~Watson,
\emph{A course of modern analysis}, Cambridge Univ.~Press (1978).
%
\bibitem{table_function}
R.~Erdelyi et al.~, \emph{Higher Trascendental Functions vol.~2},
Bateman Manuscript Project, McGraw--Hill (1953);
\\
H.~Abramowitz, I.~Stegun, \emph{Handbook of mathematical functions}, Dover
(1970).
%
\bibitem{mostepanenko_frolov}
V.~M.~Mostepanenko, V.~M.~Frolov,
Sov.~J.~Nucl.~Phys. \textbf{19}, 451 (1974).
%
\bibitem{Bogoliubov}
N.~N.~Bogoliubov, \emph{Lectures on Quantum Statistics},
Kiev: Radjanska Shkola (1949).
%
\bibitem{perelomov}
A.~M.~Perelomov, Theor.~and Math.~Phys. \textbf{16}, 852 (1973);
\\
A.~M.~Perelomov, Theor.~and Math.~Phys. \textbf{19}, 368 (1974). 
%
\bibitem{popov_marinov}
V.~S.~Popov, M.~S.~Marinov, Sov.~J.~Nucl.~Phys. \textbf{16}, 449 (1973).
%
\bibitem{pel-sor}
M.~Peloso and L.~Sorbo, JHEP \textbf{5}, 16 (2000).
%
\bibitem{bender_orszag}
C.~M.~Bender, S.~A.~Orszag,
\emph{Advanced Mathematical Methods for Scientist and Engineer}, Mc-Graw Hill (1978).
%
\bibitem{PDG}
K.~Hagiwara et al.~(Particle Data Group), Phys.~Rev.~D
\textbf{66}, 010001 (2002) [URL: http//pdg.lbl.gov].
%
\bibitem{giudice-riotto}
G.~F.~Giudice, A.~Riotto, I.~Tkachev and M.~Peloso, JHEP \textbf{8}, 14
 (1999) and references therein. 
%
\end{thebibliography}
\end{document}